\documentclass[showpacs, aps, superscriptaddress, twocolumn]{revtex4-2}
\usepackage{mathrsfs}
\usepackage{array}
\usepackage{amsmath}
\usepackage{amssymb}
\usepackage{graphicx}
\usepackage{color,xcolor}
\usepackage{booktabs}
\usepackage{amstext}
\DeclareMathOperator{\Tr}{Tr}
\usepackage{amsfonts}
\usepackage{float}
\usepackage{bm}
\usepackage{epstopdf}
\usepackage{color}

\begin{document}

\title{The differential conductance tunnel spectroscopy
in an analytical solvable\\
 two-terminal Majorana device}
\author{Chuan-Zhe Yao}
\affiliation{Department of Physics and Center for Quantum
information Science, National Cheng Kung University, Tainan 70101,
Taiwan}
\author{Wei-Min Zhang}
\email{wzhang@mail.ncku.edu.tw}
\affiliation{Department of Physics
and Center for Quantum information Science, National Cheng Kung
University, Tainan 70101, Taiwan}
\affiliation{Physics Division, National Center for Theoretical Sciences, Taipei 10617, Taiwan}

\begin{abstract}
In this paper, we investigate the non-Markovian quantum transport dynamics of a two-terminal Majorana device that is made of an asymmetric topological superconducting chain coupled to two leads. 
This asymmetric superconducting chain is analytically solvable and can be realized by a hybrid system of semiconductor nanowire coupled to 
superconductors or by 1D transverse-field Ising chains. 
In such asymmetric superconducting chains, by the change of chemical potential, its ground state undergoes a topological quantum phase transition from the topological Majorana bound state to the trivial Andreev bound state while the ground state energy remains zero. We solve the exact transient transport current and the corresponding differential conductance. The results show that the presence or absence of the interference between the left and right Majorana zero modes plays an important role on the topological phase transition of conductance. It cause the edge-localized topologically trivial states to be insulated with zero conductance, while the nonlocally distributed topologically nontrivial states always have a quantized conductance $2e^2/h$. This dramatic change associated with topological phase transition for zero-mode differential conductance at zero bias is independent of the structure of leads and the coupling strength. We also examine the finite size effect of the superconducting chain and the coherence effect between zero mode and non-zero energy modes on the differential conductance in this two-terminal Majorana device.
\end{abstract}

\maketitle
\section{Introduction}
\label{intro}
Majorana bound states have been thought to be topologically protected from quantum decoherence induced by local perturbations, and therefore have been considered as the promising candidate for potential application in quantum information and quantum computation \cite{TQC Kitaev,TQC Das Sarma, Fu15, Aasen16, Lutchyn18, Oreg20}. According to their topological characteristics, Majorana bound states are nonlocally distributed on the  
boundaries of materials, such as the ends of nanowires \cite{Kitaev chain,Lutchyn2010,Oreg2010,Alicea2010}. When the edges of the topological device are coupled to the leads, it leads to the Majorana resonance with perfect Andreev reflection \cite{resonant Andreev}, where the differential conductance $dI/dV$ at zero bias voltage (zero-bias peak) is quantized in the unit of $2e^2/h$, as theoretically predicted \cite{resonant Andreev,QZBP,review}. Recently,  experimentalists have attempted to identify Majorana bound states and to distinguish them from topologically trivial Andreev bound states \cite{QD-hybrid,QZBP exp1,QZBP exp2,Majorana nanowire1,Lutchyn18,Majorana nanowire3,hybrid exp1,hybrid exp2,hybrid exp3,Prada20,hybrid num1,hybrid num2,barrier1,barrier2,PS-Andreev1,PS-Andreev2,disorder,Large ZBP,Das Sarma2021}, based on this characteristic of the quantized zero-bias peak. 
However, due to the noises and disorders in these complicated experimental systems, identifying Majorana zero modes is experimentally challenging.

Primary detections were obtained through the tunneling spectroscopy in electrical transmission of hybrid superconductor-semiconductor nanowires \cite{QD-hybrid,QZBP exp1,QZBP exp2,Majorana nanowire1,Lutchyn18,Majorana nanowire3,hybrid exp1,hybrid exp2,hybrid exp3}. The hybrid superconductor-semiconductor nanowires are predicted to undergo a quantum phase transition from the topologically nontrivial phase to trivial phase under the external magnetic field control \cite{Lutchyn2010,Oreg2010,Alicea2010}. Up to date, the evidence for the existence of Majorana bound states has not been definitely found due to some ambiguous results that are speculated to be induced by disorder-induced subgap states in complicated material structures \cite{disorder,Large ZBP,Das Sarma2021}. Regardless the difficulties in experimentally identifying Majorana bound states, the qualitative change of conductance with topological phase transition in Majorana nanowires still has attracted considerable attentions in fundamental research and application potential. 

In this paper, we consider a two-terminal Majorana device which is an asymmetric spinless $p$-wave superconducting Kitaev chain \cite{Kitaev chain,01} coupled to two leads to study its transient transport properties. As we have shown in a recent work \cite{01}, with such an asymmetric p-wave superconducting Majorana nanowire, the wavefunction distributions of all zero and non-zero energy bogoliubon modes can be analytically solved. 
In this asymmetric superconducting chain, by controlling the chemical potential, its ground state undergoes a transition from the topologically 
nontrivial Majorana bound state to the topologically trivial Andreev bound state while the ground state energy remains at zero. From the analytical 
solution, we can unambiguously distinguish the zero energy Majorana bound state and the zero-energy Andreev bound state. 

On the other hand, the coupling between the edge state of the superconductor and the leads can be easily controlled through external electric gate or magnetic field. Thus, using the quantum Langevin equation approach \cite{PY2015,HL,Huang2020,01}, we can analytically solve the transient transport current and the differential conductance for arbitrary spectral densities of the leads at any temperature. The exact solution of the differential conductance crucially depends on the detailed coupling of the leads with the wavefunction distribution of the edge states of the superconducting chain. 
With the exact solution, we investigate how zero-bias peak is formed through non-Markovian dynamics of quantum transport. It also enable us to clarify the mechanism that causes the quantitative difference of the zero-bias conductance from $2e^2/h$ to zero between topologically non-trivial phase and topologically trivial phase. 

We further find that in the ideal case, with long superconducting chain length and large energy gap between zero-energy mode and non-zero energy band, the change of the zero bias conductance from $2e^2/h$ to zero around the phase transition point is very dramatic. The result is independent of the structure of leads and the coupling between leads and the superconducting chain. This indicates that such a device can be applied as an ideal diode 
controlled by the external filed. We also find that in the strong chain-lead coupling region 
there exists the negative differential conductance. 
The negative differential conductance has indeed been observed in molecular devices \cite{mole1,mole2,mole3,mole4,mole5} and semiconductor heterostructures \cite{hete1,hete2,hete3,hete4,hete5}. It has been applied in electronic devices such as oscillators, amplifiers, and frequency mixers. 
Meanwhile, in the situation where the energy gap between the non-zero energy state and the zero energy state is not large enough, the coherence between them will enhance zero-bias conductance peak. Such an effect could cause the zero-bias conductance peak to exceed $2e^2/h$ 
in the topologically non-trivial phase near the critical point. This is similar to the results that have been observed in recent experiments and 
are suspected to be caused by disorder-induced subgap states \cite{disorder,Large ZBP,Das Sarma2021}.

The rest of the paper is organized as follows. In Sec.~\ref{sec2}, we introduce the two-terminal Majorana device made of an asymmetric spinless $p$-wave superconducting chain coupled to leads. From the analytical solution of the ground state wavefunctions of the superconducting chain, we analyze the topological phase transition from the Majorana bound state to the Andreev bound state through the change of chemical potential. 
In Sec.~\ref{sec3}, we solve the non-Markovian decoherence dynamics of the zero-energy bogoliubov quasi-particle in the superconducting chain. The general solution of the differential conductance is also derived analytically from the transient transport current. From the general solution, we find the distinction manifested in the differential conductance associated with the difference of the wavefunction distributions between the Majorana bound states and the Andreev bound states. In Sec.~\ref{sec4}, we analyze the differential conductance and the time evolution of the zero-bias peak with different spectral widths of the leads and different lead-edge state coupling amplitudes for both the Majorana and Andreev bound states. We also investigate the effect of length of superconducting chain and the coherence between zero-energy and nonzero-energy modes. A conclusion is given in Sec.~\ref{sec6}, and the detailed derivations of the formulas are presented in Appendix.

\section{The model and its solution}
\label{sec2}
We consider an asymmetric spinless $p$-wave superconducting chain coupled to two leads. 
The total Hamiltonian of this two-terminal superconductor junction is given by
\begin{align}
H=&H_S+H_E+H_T\notag\\
=&-\sum\limits_{i=1}^{N-1}(\dfrac{t_i}{2}c^{\dagger}_i c_{i+1}+\dfrac{\Delta_i}{2} c^{\dagger}_i c^{\dagger}_{i+1} + {\rm h.c.})-\sum\limits_{i=1}^N\mu_i c^{\dagger}_i c_i\notag\\
&+\sum\limits_{\alpha=L,R}\sum\limits_k \epsilon_{\alpha k}b^{\dagger}_{\alpha k} b_{\alpha k}\notag\\
&+\sum_k (\eta_{L k}c_{1}b_{L k}^{\dagger}+\eta_{Rk}c_{N}b_{Rk}^{\dagger}+{\rm h.c.} ),
\label{H}
\end{align}
where the first two terms ($H_S$) are the Hamiltonian of the p-wave superconducting chain with superconducting gap $\Delta_i$, 
hopping amplitude $t_i$, and electron chemical potential $\mu_i$ at site $i$, and $c^{\dagger}_i$ and $c_i$ are the corresponding creation 
and annihilation 
operators of electrons. The third term ($H_E$) is the Hamiltonian of the two leads, and $b^{\dagger}_{\alpha k}$ and 
$b_{\alpha k}$ are the creation and annihilation operators of electrons of the $k$th level in lead $\alpha$ with the corresponding eigenenergy spectra
$\epsilon_{\alpha k}$. The last term ($H_T$) describes the tunnellings between the superconductor and the leads, where $\eta_{Lk}$ 
($\eta_{Rk}$) is the coupling amplitude between the left lead and the first site of the superconducting chain (the right lead 
and the last site of the superconducting chain).

The asymmetric superconducting chain was inspired from the asymmetric transverse-field Ising model we introduced in a recent work \cite{01}, that is, set the electron 
chemical potential (the on-site energy) on the last site $\mu_N=0$ in the superconducting chain. Also for simplicity, we set $t_i=t$, $\Delta_i=|\Delta|e^{i\theta}$ and $\mu_i=\mu\ (i=1,2,\cdots,N-1)$ with $t=|\Delta|$.
Such an asymmetric setting makes the Hamiltonian of the superconducting chain analytically solvable and allows the ground state undergoes a transition from the Majorana 
bound state to the Andreev bound state when we vary the chemical potential $\mu$, while the ground state energy remains at zero. 

To be explicit, we apply the Bogoliubov transformation 
\begin{subequations}
\label{B0}
\begin{align}
a_{j}&=\sum\limits_{i=1}^{N}(u_{ji}c_{i}+v_{ji}c_{i}^{\dagger})\\
a_{j}^{\dagger}&=\sum\limits_{i=1}^{N}(v_{ji}^{*}c_{i}+u_{ji}^{*}c_{i}^{\dagger}),
\end{align}
\end{subequations}
to the superconducting chain Hamiltonian, its diagonalized form is given by
\begin{align}
H_{S}=\sum\limits_{j}\varepsilon_{j}(a_{j}^{\dagger}a_{j}-a_{j}a_{j}^{\dagger}).
\label{B1}
\end{align}
Here $a^{\dagger}_j$ and $a_{j}$ are creation and annihilation operators of Bogoliubov quasi-particles (bogoliubons) with the spectrum
\begin{align}
\varepsilon_{j}=\left\{\begin{array}{ll}
\dfrac{|\Delta|}{2}\sqrt{1+\lambda^{2}-2\lambda\cos\frac{j\pi}{N}}, & j=1,2,\ldots,N-1\\
0, &  j=0
\end{array} \right. ,
\label{e}
\end{align}
where $\lambda=\mu/|\Delta|$. The corresponding wavefunctions for the non-zero-energy bogoliubons can be analytically solved
\begin{subequations}
\label{kmode}
\begin{align}
u_{ji}&=\mathcal{N}_{j}e^{-i\frac{\theta}{2}}\bigg\{\!\frac{-|\Delta|}{\varepsilon_j}\sin\!\bigg[\dfrac{(i\!-\!1)j\pi}{N}\bigg]\!+\!\bigg(\!1\!-\!\dfrac{|\Delta|\lambda}{\varepsilon_{j}}\!\bigg)\sin \dfrac{ij\pi}{N}\!\bigg\}, \\
v_{ji}&=\mathcal{N}_{j}e^{i\frac{\theta}{2}}\bigg\{\!\frac{-|\Delta|}{\varepsilon_{j}}\sin\!\bigg[\dfrac{(i\!-\!1)j\pi}{N}\bigg]\!-\!\bigg(\!1\!+\!\dfrac{|\Delta|\lambda}{\varepsilon_{j}}\!\bigg)\sin \dfrac{ij\pi}{N}\!\bigg\} .
\end{align}
\end{subequations}
The wavefunction of the zero-energy bogoliubon is 
\begin{subequations}
\label{zero}
\begin{align}
u&_{0 i}=\left\{\begin{array}{l}
e^{-i\frac{\theta}{2}}\mathcal{N}_{0}(-\lambda)^{i-1}\hfill i<N\\
e^{-i\frac{\theta}{2}}[\mathcal{N}_{0}(-\lambda)^{N-1}+1/2]\quad\hfill i=N
\end{array} \right.\\
v&_{0 i}=\left\{\begin{array}{l}
e^{-i\frac{\theta}{2}}\mathcal{N}_{0}(-\lambda)^{i-1}\hfill i<N\\
e^{-i\frac{\theta}{2}}[\mathcal{N}_{0}(-\lambda)^{N-1}-1/2]\quad\hfill i=N
\end{array} \right. .
\end{align}
\end{subequations}
The normalized constants $\mathcal{N}_{j},\mathcal{N}_{0}$ are given by
\begin{subequations}
\begin{align}
\mathcal{N}_{j}&=\bigg\{N\Big[1+2\dfrac{|\Delta|^{2}}{\epsilon^{2}_{j}}(1+\cos \dfrac{j\pi}{N})\Big]\bigg\}^{-1/2}\\
\mathcal{N}_{0}&=\dfrac{1}{2}\bigg(\dfrac{1-\lambda^{2N}}{1-\lambda^{2}}\bigg)^{-1/2} .
\end{align}
\end{subequations}

\begin{figure}[t]
\includegraphics[scale=0.37]{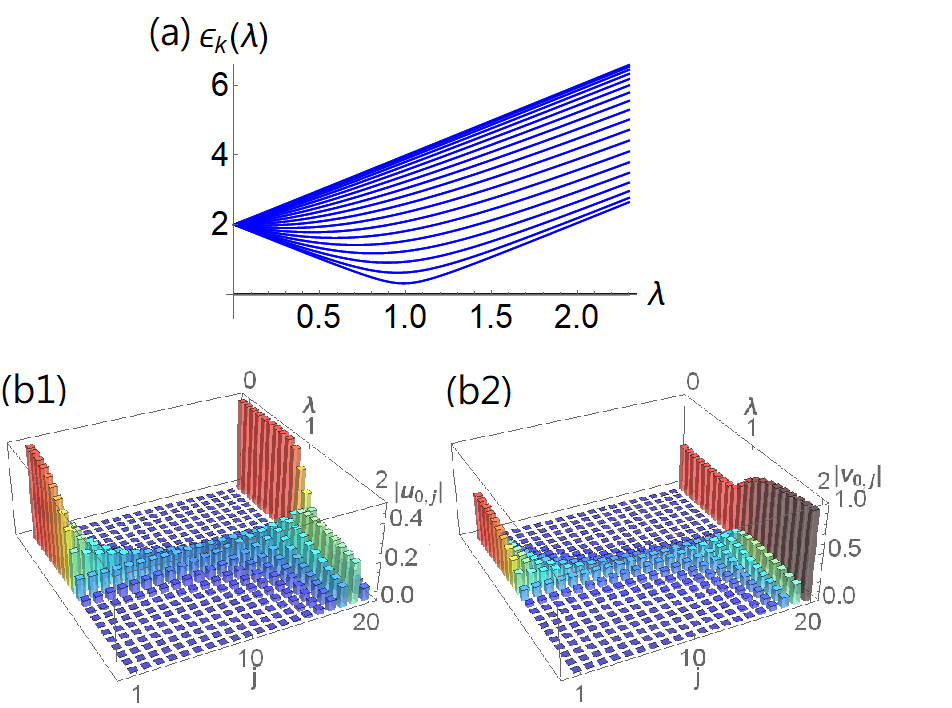}
\caption{(Colour online) (a) The spectrum of the asymmetric superconducting chain model in Eq.~(\ref{H}) with $N=20$. 
(b) The wavefunction distribution of the zero energy ground state (zero modes), 
$|u_{0,i}|$ (top) and $|v_{0,i}|$ (bottom) varying from the topologically nontrivial phase $\lambda<1$ 
to the topologically trivial phase $\lambda>1$.} \label{fig1}
\end{figure}

Note that the zero-energy bogoliubon states ($j=0$) are twofold degenerate states, with the particle number $a^{\dagger}_{0} a_{0}=0$ and $1$, 
respectively. These states can be described by the left and right Majorana operators $\gamma_{L0}=a_{0}+a^{\dagger}_{0},\ \gamma_{R0}=-i(a_{0}-a^{\dagger}_{0})$. More importantly, the superconducting chain undergoes a topological phase transition when $\lambda$ across the 
critical point $\lambda_c=1$, which can be clearly seen in the ground state wavefunction distributions, as shown in Fig.~\ref{fig1}(b). 
In the topologically nontrivial phase $\lambda<1$, the two Majorana zero modes are nonlocally separated in the two sides of the superconducting chain, 
which forms two Majorana bound state. As $\lambda$ gets larger and larger, the left Majorana zero mode spreads into other sites, 
while the right Majorana zero mode remains unchanged due to the asymmetric setting $\mu_{N}=0$. At the critical point $\lambda_{c}=1$, 
the left Majorana zero mode is uniformly distributed in the all sites of the chain. When $\lambda>1$, the wavefunction of the left Majorana zero 
mode distributes more on the right-hand side than the left-hand side. With continuously increasing $\lambda$, both the two Majorana zero 
modes $\gamma_{L0},\gamma_{R0}$ will eventually condense to the last site $N$, and therefore the zero-energy states still exist but no longer 
have the topologically nonlocal property. In other words, they become topologically trivial zero-energy Andreev bound states for $\lambda>1$.

Now, in terms of bogoliubons of Eq.~(\ref{B0}), the total Hamiltonian of the two-terminal device can be re-expressed as
\begin{align}
H = & \sum\limits_{j}\varepsilon_{j}(a_{j}^{\dagger}a_{j}-a_{j}a_{j}^{\dagger})
+ \!\!\! \sum\limits_{\alpha=L,R} \!\! \sum\limits_k \epsilon_{\alpha k}b^{\dagger}_{\alpha k} b_{\alpha k} \notag \\
&+ \sum\limits_{j,k} \big[ \eta_{Lk} (\kappa_{Lj}a_{j}+\kappa^{\prime}_{Lj}a^{\dagger}_{j}) b_{L k}^{\dagger}\notag\\
&~~~~~~~~~~+\eta_{Rk}(\kappa_{Rj}a_{j}+\kappa^{\prime}_{Rj}a^{\dagger}_{j})b_{Rk}^{\dagger}+{\rm h.c.}\big] 
\label{Hb}
\end{align}
where
\begin{subequations}
\label{tunnelingc}
\begin{align}
&\kappa_{Lj}=\left\{\begin{array}{ll}
e^{i\frac{\theta}{2}}\sqrt{\frac{1}{2N}}(1+\frac{|\Delta|\lambda}{\epsilon_{j}})\sin\frac{j\pi}{N} , & j =1, 2, \cdots, N-1 \\
e^{i\frac{\theta}{2}}(\sum_{i=0}^{N-1}\lambda^{2i})^{-1/2}/2 , & j=0
\end{array} \right.\\
&\kappa_{Lj}^{\prime}=\left\{\begin{array}{ll}
e^{i\frac{\theta}{2}}\sqrt{\frac{1}{2N}}(1-\frac{|\Delta|\lambda}{\epsilon_{j}})\sin\frac{j\pi}{N} , & j =1, 2, \cdots, N-1 \\
e^{i\frac{\theta}{2}}(\sum_{i=0}^{N-1}\lambda^{2i})^{-1/2}/2 , & j=0
\end{array} \right.\\
&\kappa_{Rj}=\left\{\begin{array}{ll}
e^{i\frac{\theta}{2}}\frac{|\Delta|(-1)^{j}}{2\epsilon_{j}\sqrt{N}} \sin\frac{j\pi}{N},~~~~~~~~~~~ j =1, 2, \cdots, N-1 \\
e^{i\frac{\theta}{2}}[(-\lambda)^{N-1}(\sum_{i=0}^{N-1}\lambda^{2i})^{-1/2}+1]/2 ,~~ j=0
\end{array} \right.\\
&\kappa_{Rj}^{\prime}=\left\{\begin{array}{ll}
-e^{i\frac{\theta}{2}}\frac{|\Delta|(-1)^{j}}{2\epsilon_{j}\sqrt{N}}\sin\frac{j\pi}{N} ,~~~~~~~~~ j =1, 2, \cdots, N-1 \\
e^{i\frac{\theta}{2}}[(-\lambda)^{N-1}(\sum_{i=0}^{N-1}\lambda^{2i})^{-1/2}-1]/2 ,~~ j=0
\end{array} \right. ,
\end{align}
\end{subequations}
are the inverse Bogoliubov transformation coefficients of Eq.~(\ref{B0}) for the electron operator $c_1$ and $c_N$.
In the Bogoliubov basis, the tunneling Hamiltonian in Eq.~(\ref{Hb}) shows all kind of tunneling channels between the 
superconducting chain and 
leads through the zero and non-zero energy states of the superconducting chain.

In the large $N$ limit ($N\rightarrow\infty$), other tunneling amplitudes in Eq.~(\ref{tunnelingc}) become negligibly small
except for these between the zero-energy bogoliubon and two leads. They are simply reduced to
\begin{subequations}
\label{coupling}
\begin{align}
&\kappa_{L0}=\kappa_{L0}^{\prime}=\left\{\begin{array}{ll}
e^{i\frac{\theta}{2}}\sqrt{1-\lambda^2}/2\ &\lambda<1\\
0 &\lambda\geq1
\end{array}\right.\\
&\kappa_{R0}=\left\{\begin{array}{ll}
e^{i\frac{\theta}{2}}/2 &\lambda<1\\
e^{i\frac{\theta}{2}}[\sqrt{1-\lambda^{-2}}+1]/2\ &\lambda\geq1
\end{array}\right.\\
&\kappa_{R0}^{\prime}=\left\{\begin{array}{ll}
-e^{i\frac{\theta}{2}}/2 &\lambda<1\\
e^{i\frac{\theta}{2}}[\sqrt{1-\lambda^{-2}}-1]/2\ &\lambda\geq1
\end{array} \right. .
\end{align}
\end{subequations}
It shows that in the topologically nontrivial phase $\lambda<1$, Eq.~(\ref{Hb}) becomes
\begin{align}
H = & \sum\limits_{j}\varepsilon_{j}(a_{j}^{\dagger}a_{j}-a_{j}a_{j}^{\dagger})
+\!\! \sum\limits_{\alpha=L,R}\sum\limits_k \epsilon_{\alpha k}b^{\dagger}_{\alpha k} b_{\alpha k} \notag \\
&+ \!\sum\limits_{k} \big[e^{i\frac{\theta}{2}}( \eta_{Lk} \sqrt{1\!-\!\lambda^2}\gamma_{L0} b_{L k}^{\dagger}
+i\eta_{Rk}\gamma_{R0}b_{Rk}^{\dagger})\!+\!{\rm h.c.}\big] .
\label{Hm}
\end{align}
This Hamiltonian simply describes the tunneling between the left (right) lead with the left (right) Majorana zero mode. In contrast, the Hamiltonian in the topologically trivial phase $\lambda>1$ is reduced to
\begin{align}
H = & \sum\limits_{j}\varepsilon_{j}(a_{j}^{\dagger}a_{j}-a_{j}a_{j}^{\dagger})
+\sum\limits_{\alpha=L,R}\sum\limits_k \epsilon_{\alpha k}b^{\dagger}_{\alpha k} b_{\alpha k} \notag \\
&+ \sum\limits_{k} \eta_{Rk} \big[e^{i\frac{\theta}{2}}(\sqrt{1-\lambda^{-2}}\gamma_{L0} +i\gamma_{R0})b_{Rk}^{\dagger}+{\rm h.c.}\big] .
\label{Ha}
\end{align}
It shows that [also see from Fig.~\ref{fig1}(b)] the left Majorana zero mode wavefunction distributing to the right side 
of the chain when $\lambda>1$, and locally overlaps with the right Majorana zero mode eventually. That is, the original zero 
energy Majorana bound state for $\lambda <1$ becomes the topological trivial zero-energy Andreev bound state for $\lambda >1$. Therefore, Eq.~(\ref{Ha}) 
contains only the tunneling through the zero-energy Andreev bound state. The change of the system-lead coupling from Eq.~(\ref{Hm}) to
Eq.~(\ref{Ha}) with increasing $\lambda$ provides a unambiguous picture for exploring the difference of quantum transport via 
Majorana bound state and Andreev bound state in this two-terminal device.

\section{the exact differential conductance}
\label{sec3}

\subsection{General formulation of quantum transport}
The quantized zero-bias peak of the differential conductance $dI/dV$ has been considered as the signature of Majorana bound state in the hybrid system of the superconductor-semiconductor nanowire junctions, while the Andreev bound state is thought to have no such characteristic \cite{resonant Andreev,QZBP}. In order to understand more specifically the possible physical origin for this difference, we calculate the differential conductance which is defined as $dI_\alpha(t)/dV$, and $\alpha=L,R$, where $I_\alpha(t)$ is the electron current flowing from the lead $\alpha$ into the asymmetric superconducting chain. The transient transport current flowing through all possible channels in the superconducting chain in terms of the generalized nonequilibrium Green functions \cite{PY2018}
\begin{align}
I_{\alpha}(t) & \!=\!\frac{2e}{\hbar}\! \operatorname{Re}\!\bigg[\! \int_{t_{0}}^{t} \!\!\! d\tau \! \Tr\! \Big[\boldsymbol{\mathcal{\tilde{G}}_{\alpha}}(t,\!\tau) 
\boldsymbol{U^{\dagger}}(t,\!\tau)
\!-\!\boldsymbol{\mathcal{G}_{\alpha}}(t,\!\tau)\boldsymbol{V}(\tau,t) \Big]\!\bigg] ,
\label{I}
\end{align}
where the functions $\boldsymbol{U}(\tau,t_0)$ and $\boldsymbol{V}(\tau,t)$ satisfy 
\begin{subequations}
\begin{align}
&\dfrac{d}{dt}\boldsymbol{U}(\tau,t_{0})\!+\!\dfrac{i}{\hbar}\!\begin{pmatrix}
\boldsymbol{\epsilon}&0\\
0&-\boldsymbol{\epsilon}
\end{pmatrix} \boldsymbol{U}(\tau,t_{0})\notag\\
&~~~~~~~~~~~~~+\int_{t_{0}}^{\tau} \boldsymbol{G}(t,\tau_1)\boldsymbol{U}(\tau_1,t_{0})d\tau_1=0, \label{U}\\
&\boldsymbol{V}(\tau, t)= \!\!\int_{t_{0}}^{\tau} \!\!\!\! d\tau_{1} \!\! \int_{t_{0}}^{t}\! \!\! d\tau_{2} \boldsymbol{U}(\tau,\tau_{1})
\boldsymbol{\tilde{G}}(\tau_{1},\tau_{2}) \boldsymbol{U^{\dagger}}(t,\tau_{2}),
\label{V}
\end{align}
\end{subequations}
subject to the boundary conditions $\boldsymbol{U}(t_0,t_0)=\boldsymbol{I}$ and $\boldsymbol{V}(t_0,t)=\boldsymbol{0}$. The detailed derivation 
of the above results in terms of the exact non-Markovian dynamics of bogoliubons in the superconducting chain is summarized in the Appendix.

The integral kernel $\boldsymbol{G}(t,\tau)$ in Eq.~(\ref{U}) characterizes all the non-Markovian memory effects arisen from back-reactions of electron transport between the 
superconducting chain and the two leads through all zero- and nonzero-energy bogoliubon state channels. It can be expressed as
\begin{align}
\boldsymbol{G}(t,\tau)=&\frac{1}{\hbar^2} \sum\limits_{\alpha}\int\dfrac{d\omega}{2\pi}e^{-i\omega(t-\tau)}\notag\\
&\times\!\!\left(\!\!\begin{array}{cc}
\boldsymbol{J}_{\alpha}(\omega)+\boldsymbol{\hat{J}}_{\alpha}(-\omega) 
&\boldsymbol{\bar{J}}_{\alpha}(\omega)+\boldsymbol{\bar{J}}_{\alpha}(-\omega)\\
\boldsymbol{\bar{J}}_{\alpha}(\omega)+\boldsymbol{\bar{J}}_{\alpha}(-\omega)
&\boldsymbol{J}_{\alpha}(-\omega)+\boldsymbol{\hat{J}}_{\alpha}(\omega)
\end{array}\!\!\right) .
\label{G0}
\end{align}
The corresponding spectral density matrix elements,
\begin{subequations}
\label{efsd}
\begin{align}
&J_{\alpha ij}(\omega)=|\kappa_{\alpha i}\kappa_{\alpha j}|J_\alpha(\omega), \\
&\hat{J}_{\alpha ij}(\omega)=|\kappa'_{\alpha i}\kappa'_{\alpha j}| J_\alpha(\omega), \\
&\bar{J}_{\alpha ij}(\omega)=|\kappa_{\alpha i}\kappa'_{\alpha j}| J_\alpha(\omega),
\end{align}
\end{subequations}
represent the normal tunnelings, the Andreev reflections and the mixing process, respectively.
In Eq.~(\ref{efsd}),  $J_\alpha(\omega)=2\pi \sum_{k}|\eta_{\alpha k}|^2\delta(\omega-\epsilon _{\alpha k}/\hbar)$ 
is the spectral density of lead $\alpha$. Note that the inverse Bogoliubov transformation coefficients 
$\kappa_{\alpha i}, \kappa'_{\alpha j}$ given by Eq.~(\ref{tunnelingc}) are different for the topological nontrivial phase ($\lambda <1$) and topological trivial phase ($\lambda >1$), so does the corresponding transient transport currents. 
The integral kernel $\boldsymbol{\tilde{G}}(t_1,t_2)$ in Eq.~(\ref{V}) is given by
\begin{align}
&\boldsymbol{\tilde{G}}(t_1,t_2)\!=\!\! \sum\limits_{\alpha} \!\!\int \!\!\dfrac{d\omega}{2\pi}e^{-i\omega(t_1-t_2)}\bigg[f_{\alpha}(\omega)
\!\left(\!\begin{array}{cc}
\boldsymbol{J}_{\alpha}(\omega) 
&\boldsymbol{\bar{J}}_{\alpha}(\omega)\\
\boldsymbol{\bar{J}}_{\alpha}(\omega)
&\boldsymbol{\hat{J}}_{\alpha}(\omega)
\end{array}\!\right)\notag\\
&~~~~~~~~~~~~~~~~+[1\!-\!f_{\alpha}(-\omega)]\!\left(\!\begin{array}{cc}
\boldsymbol{\hat{J}}_{\alpha}(-\omega) 
&\boldsymbol{\bar{J}}_{\alpha}(-\omega)\\
\boldsymbol{\bar{J}}_{\alpha}(-\omega)
&\boldsymbol{J}_{\alpha}(-\omega)
\end{array}\!\right)\!\bigg],
\end{align}
which is associated with the initial thermal operators 
$\{b_{\alpha k}(t_{0}),b^{\dagger}_{\alpha k}(t_{0})\}$  of the lead $\alpha=L,R$. It is 
given by the Fermi-Dirac distribution of electrons in leads at initial time $t_0$: $f_\alpha (\omega_k)
= \langle b^{\dagger}_{\alpha k}(t_{0})b_{\alpha k}(t_{0}) \rangle = [e^{\beta(\omega-\mu_\alpha)}+1]^{-1}$, 
and $\mu_{L,R}=E_F\pm eV$ where $E_F$ is the Fermi level and $V$ is the bias voltage anti-symmetrically applied to the two leads. The other two-time system-lead correlation functions in Eq.~(\ref{I}) are defined by 
\begin{subequations}
\begin{align}
&\boldsymbol{\mathcal{G}_{\alpha}}(t,\tau)=\frac{1}{\hbar^2}\int\dfrac{d\omega}{2\pi}e^{-i\omega(t-\tau)}\notag\\
&~~~~~~~~~~~\times\!\left(\!\!\begin{array}{cc}
\boldsymbol{J}_{\alpha}(\omega)-\boldsymbol{\hat{J}}_{\alpha}(-\omega) 
&\boldsymbol{\bar{J}}_{\alpha}(\omega)-\boldsymbol{\bar{J}}_{\alpha}(-\omega)\\
\boldsymbol{\bar{J}}_{\alpha}(\omega)-\boldsymbol{\bar{J}}_{\alpha}(-\omega)
&\boldsymbol{J}_{\alpha}(-\omega)-\boldsymbol{\hat{J}}_{\alpha}(\omega)
\end{array}\!\!\right)  \\
&\boldsymbol{\mathcal{\tilde{G}}_{\alpha}}(t,\tau)
=\!\! \int \!\! \dfrac{d\omega}{2\pi}e^{-i\omega(t-\tau)}\bigg[f_{\alpha}(\omega)\!\left(\!\begin{array}{cc}
\boldsymbol{J}_{\alpha}(\omega) 
&\boldsymbol{\bar{J}}_{\alpha}(\omega)\\
\boldsymbol{\bar{J}}_{\alpha}(\omega)
&\boldsymbol{\hat{J}}_{\alpha}(\omega)
\end{array}\!\right)\notag \\
&~~~~~~~~~~~~~~~-\![1\!-\!f_{\alpha}(-\omega)]\!\left(\!\begin{array}{cc}
\boldsymbol{\hat{J}}_{\alpha}(-\omega) 
&\boldsymbol{\bar{J}}_{\alpha}(-\omega)\\
\boldsymbol{\bar{J}}_{\alpha}(-\omega)
&\boldsymbol{J}_{\alpha}(-\omega)
\end{array}\!\right)\!\bigg]
\end{align}
\end{subequations}
Note that in Eq.~(\ref{I}), only $\boldsymbol{\mathcal{\tilde{G}}_{\alpha}}(t,\tau)$ and $\boldsymbol{V}(t,\tau)$ (through 
$\boldsymbol{\tilde{G}}(t,\tau)$) are related to the bias voltage $V$, and thus one can find the differential 
conductance explicitly by taking the derivative of these two functions in Eq.~(\ref{I}) with respect to the bias voltage $V$. 

Due to the particle-hole symmetry, the spectral density is symmetric, $\boldsymbol{J}(\omega)=\boldsymbol{J}(-\omega)$. We also assume that the leads are initially at zero temperature $\beta\rightarrow\infty$. Taking the steady-state limit $t\rightarrow\infty$ (see Eq.~(\ref{condT}) and Eq.~(\ref{cond})), the differential conductance can be reduced to a simple form
\begin{align}
\dfrac{dI_\alpha}{dV}=&\dfrac{e^2}{h}\operatorname{Re}\, \Tr\bigg[[\boldsymbol{\mathbb{J}_{\alpha}}(\tilde{V})\!+\!\boldsymbol{\hat{\mathbb{J}}_{\alpha}}(\tilde{V})]\boldsymbol{\tilde{U}}(\tilde{V})\notag\\
&\pm [\boldsymbol{\mathbb{J}_{\alpha}}(\omega)\!-\!\boldsymbol{\hat{\mathbb{J}}_{\alpha}}(\omega)]\boldsymbol{\tilde{U}}(\tilde{V})[\boldsymbol{\mathbb{J}_{M}}(\tilde{V})\!-\!\boldsymbol{\hat{\mathbb{J}}_{M}}(\tilde{V})]\boldsymbol{\tilde{U}^\dagger}(\tilde{V})
\bigg],
\label{DC}
\end{align}
where the sign $\pm$ is determined by the direction of bias voltage initially applied to the left and right leads (we take $+V$ for left lead and $-V$ for right lead in the calculation given in Appendix). The function $\boldsymbol{\tilde{U}}(\tilde{V})$ is the Laplace transformation of $\boldsymbol{U}(t,t_0)$ in the frequency $\tilde{V}=eV/\hbar$ domain, as shown in Eq.~(\ref{LaplaceU}). 
The spectral density matrices are given by
\begin{subequations}
\label{spectral_pm}
\begin{align}
&\boldsymbol{\mathbb{J}_\alpha}(\omega)=\left(\begin{array}{cc}
\boldsymbol{J}_{\alpha}(\omega) 
&\boldsymbol{\bar{J}}_{\alpha}(\omega)\\
\boldsymbol{\bar{J}}_{\alpha}(\omega)
&\boldsymbol{\hat{J}}_{\alpha}(\omega)
\end{array}\right)\\
&\boldsymbol{\hat{\mathbb{J}}_\alpha}(\omega)=\left(\begin{array}{cc}
\boldsymbol{\hat{J}}_{\alpha}(\omega) 
&\boldsymbol{\bar{J}}_{\alpha}(\omega)\\
\boldsymbol{\bar{J}}_{\alpha}(\omega)
&\boldsymbol{J}_{\alpha}(\omega)
\end{array}\right),
\end{align}
\end{subequations}
where $\boldsymbol{\mathbb{J}_{M}}=\boldsymbol{\mathbb{J}_{L}}-\boldsymbol{\mathbb{J}_{R}}$ , $\boldsymbol{\hat{\mathbb{J}}_{M}}=\boldsymbol{\hat{\mathbb{J}}_{L}}-\boldsymbol{\hat{\mathbb{J}}_{R}}$. Equation~(\ref{DC}) shows that the differential conductance sensitively depends on the spectral densities which are proportional to the inverse Bogoliubov transformation coefficients $\boldsymbol{\kappa}_{\alpha}$ and $\boldsymbol{\kappa}'_{\alpha}$, see Eq.~(\ref{efsd}). Furthermore, recall that these coupling amplitudes are determined by bogoliubon wavefunctions and vary when the superconductor chain undergoes a transition from the topological Majorana bound state to the trivial Andreev bound state, as shown in Eq.~(\ref{tunnelingc}). Therefore, the differential conductance can reveal unambiguously the topological quantum phase transition from the Majorana bound state to the Andreev bound state in this two-terminal quantum device.

\subsection{Differential conductance through the zero energy channel}
Except for the critical regime (near the critical point $\lambda_c=1$), there is a relatively large energy gap between the zero-energy mode and the non-zero bulk band. We can reasonably assume that the zero energy state has negligible coherence with other excited states if the applying bias is much smaller than the superconducting gap. Thus we can focus on the transport current through the zero energy state in the superconducting chain, and the $N\times N$ matrix of the spectral densities will be reduced to a $2\times 2$ matrix. The corresponding electron transport currents are denoted as $I_{0\alpha}$ for $\alpha=L, R$. 

\subsubsection{Differential conductance with the left lead}
We first focus on the current flow from the left lead into the zero-energy bogoliubon in the superconducting chain. From Eq.~(\ref{DC}) and Eq.~(\ref{U0}), one can explicitly find the solution of the corresponding differential conductance, which can be expressed as
\begin{align}
\dfrac{dI_{0L}}{dV}=&\dfrac{2e^2}{h}\bigg[\dfrac{\mathcal{J}_{+}(\tilde{V})}{[\tilde{V}\!+\!\delta\omega_+(\tilde{V})]^2\!+\!\mathcal{J}_{+}^2(\tilde{V})} \Big[\mathcal{J}_{+L}(\tilde{V})\!-\!{\mathcal{J}}_{+-L}(\tilde{V}) \Big]\notag\\
&+\!\dfrac{\mathcal{J}_{-}(\tilde{V})}{[\tilde{V}\!+\!\delta\omega_-(\tilde{V})]^2\!+\!\mathcal{J}_{-}^2(\tilde{V})}
 \Big[\mathcal{J}_{-L}(\tilde{V})\!-\!{\mathcal{J}}_{+-L}(\tilde{V}) \Big] \bigg] .
\label{steadyL}
\end{align}
Here the energy shift $\delta\omega_{\pm}(s)=\mathcal{P}\int\frac{d\omega}{\pi}\frac{\mathcal{J}_{\pm}(\omega)}{s-\omega}$, and ${\cal P}$ denotes the principal values of the integrals. We also introduced the effective spectral densities $\mathcal{J}_{\pm\alpha}(\omega)=|\kappa_{\alpha 0}\pm\kappa_{\alpha 0}^\prime|^2J_\alpha(\omega)$, and $\mathcal{J}_{\pm}(\omega)=\mathcal{J}_{\pm L}(\omega)+\mathcal{J}_{\pm R}(\omega)$. The effective spectral density $\mathcal{J}_{+\alpha}$ relates to the coupling between lead $\alpha$ and the left Majorana zero mode, and $\mathcal{J}_{-\alpha}$ relates to the coupling between lead $\alpha$ and the right Majorana zero mode.
Moreover, ${\mathcal{J}}_{+-L}(\tilde{V})$ is a crossing spectral density defined by
\begin{align}
{\mathcal{J}}_{+-L}(\tilde{V})=&2\left(\!\sqrt{\mathcal{J}_{+L}(\tilde{V})\mathcal{J}_{-L}(\tilde{V})}-\!\sqrt{\mathcal{J}_{+R}(\tilde{V})\mathcal{J}_{-R}(\tilde{V})}\right)\notag\\
&\times\dfrac{\sqrt{\mathcal{J}_{+L}(\tilde{V})\mathcal{J}_{-L}(\tilde{V})}}{\mathcal{J}_{+}(\tilde{V})\!+\!\mathcal{J}_{-}(\tilde{V})}.
\end{align}
The terms that contain ${\mathcal{J}}_{+-L}(\tilde{V})$ in Eq.~(\ref{steadyL}) describe the interference between the current transport through the channels of the left and right Mojorana zero modes, induced by the indirect coupling between the two Majorana zero modes through the lead.
Furthermore, Eq.~(\ref{coupling}) shows that the inverse Bogoliubov transformation coefficients $\kappa_{0L}$ and $\kappa^\prime_{0L}$ are always the same so that ${\cal J}_{-L}(\omega)=0$. This implies that the left lead only couples with the left Majorana zero mode and the right Majorana zero mode has no contribution to the  left differential conductance. More importantly, it also shows that the interference terms are vanished because ${\mathcal{J}}_{+-L}(\omega)=0$ due to ${\cal J}_{-L}(\omega)=0$.  Thus, the left differential conductance Eq.~(\ref{steadyL}) is further reduced to 
\begin{align}
\dfrac{dI_{0L}}{dV}=&\dfrac{2e^2}{h}\dfrac{\mathcal{J}_{+}(\tilde{V})\mathcal{J}_{+L}(\tilde{V})}{[\tilde{V}\!+\!\delta\omega_+(\tilde{V})]^2\!+\!\mathcal{J}_{+}^2(\tilde{V})} .
\label{steadyL0}
\end{align}

\begin{figure}[t]
\includegraphics[scale=0.2]{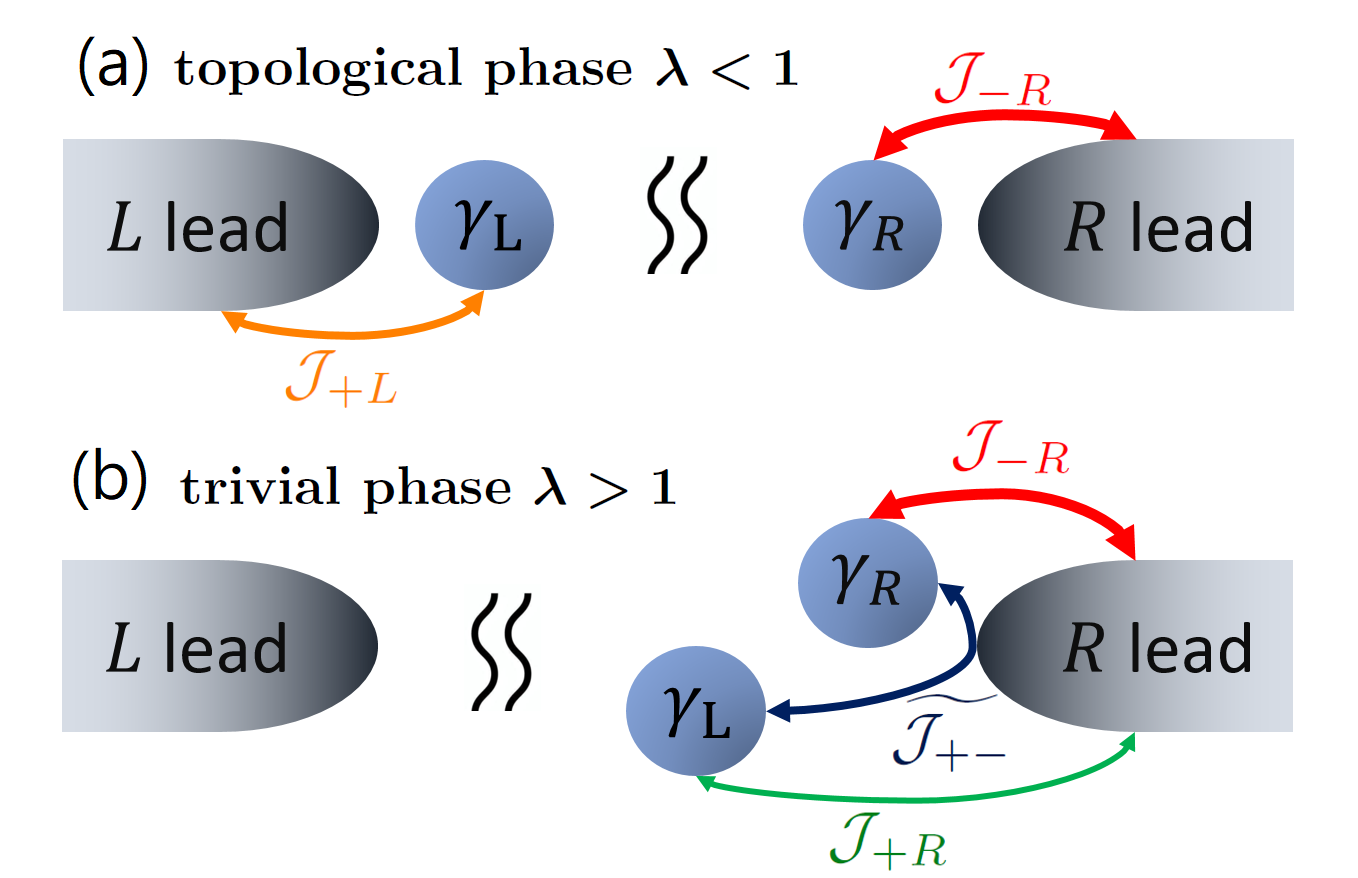}
\caption{(Colour online) A schematic plot of the different electron transport paths between the leads and the left and right Majorana zero modes with the corresponding effective spectral densities: (a) in the topologically nontrivial phase and (b) in the topologically trivial phase.} \label{fig2}
\end{figure}

In the topologically nontrivial phase ($\lambda<1$), we have $\mathcal{J}_{+R}(\omega)=0$ and $\mathcal{J}_{+L}(\omega) =\sqrt{1-\lambda^2}J_L(\omega)=\mathcal{J}_{+}(\omega)$, as shown in Fig.~\ref{fig2}(a). The left differential conductance for $\lambda<1$ is simply reduced to 
\begin{align}
\dfrac{dI_{0L}}{dV}=\dfrac{2e^2}{h}\bigg[& \dfrac{(1-\lambda^2)J_L^2(\tilde{V})}{[\tilde{V}\!+\!\delta\omega_{+L}(\tilde{V})]^2\!+\!(1-\lambda^2)J_L^2(\tilde{V})}
\bigg],
\label{left cond0}
\end{align}
which reproduces the result in Ref.~\cite{Huang2020} as a special case with the symmetric spectral density. Note that $\delta\omega_{+L}(\tilde{V})$ is an odd function for the symmetric spectral density, so it vanishes at zero bias. Consequently, the zero-bias differential conductance has a quantized value $2e^2/h$, independent of the structure of spectral density $J_L(\omega)$.
In the topologically trivial phase $(\lambda >1)$, both the coefficients $\kappa_{0L}$ and $\kappa^\prime_{0L}$ vanish, and thus $\mathcal{J}_{+L}(\omega)=0$. This corresponds to a trivial case that the left lead decouples completely from the left Majorana zero mode, as shown in Fig.~\ref{fig2}(b). This is because for $\lambda >1$, the left Majorana zero mode moves to the right side of the superconducting chain, see Fig.~\ref{fig1}(b). Thus the left differential conductance for $\lambda>1$ is always zero,
\begin{align}
\dfrac{dI_{0L}}{dV}=0.
\end{align}

\subsubsection{Differential conductance with the right lead}

The differential conductance associated with the transport current flowing into the right lead is given by
\begin{align}
\dfrac{dI_{0R}}{d(-V)}\!=& \, \!\dfrac{2e^2}{h}\!\bigg[\!\dfrac{\mathcal{J}_{+}(\tilde{V})}{[\tilde{V}\!+\!\delta\omega_+(\tilde{V})]^2\!+\!\mathcal{J}_{+}^2(\tilde{V})} \Big[\!\mathcal{J}_{+R}(\!\tilde{V}\!)\!+\!{\mathcal{J}}_{+-R}(\!\tilde{V}\!) \!\Big]\notag\\
&+\!\dfrac{\mathcal{J}_{-}(\tilde{V})}{[\tilde{V}\!+\!\delta\omega_-(\tilde{V})]^2\!+\!\mathcal{J}_{-}^2(\tilde{V})}
 \Big[\mathcal{J}_{-R}(\tilde{V})\!+\!{\mathcal{J}}_{+-R}(\tilde{V}) \Big] \bigg] ,
\label{steadyR}
\end{align}
The derivative of the current with the negative bias voltage is because the bias is applied anti-symmetrically to the two leads. The crossing 
spectral density is given by
\begin{align}
{\mathcal{J}}_{+-R}(\tilde{V})=&2\left(\!\sqrt{\mathcal{J}_{+L}(\tilde{V})\mathcal{J}_{-L}(\tilde{V})}-\!\sqrt{\mathcal{J}_{+R}(\tilde{V})\mathcal{J}_{-R}(\tilde{V})}\right)\notag\\
&\times\dfrac{\sqrt{\mathcal{J}_{+R}(\tilde{V})\mathcal{J}_{-R}(\tilde{V})}}{\mathcal{J}_{+}(\tilde{V})\!+\!\mathcal{J}_{-}(\tilde{V})}.
\end{align}
Notice that the inverse Bogoliubov transformation coefficients of the left Majorana zero modes  $\kappa_{0L}=\kappa^\prime_{0L}$ so that 
$\mathcal{J}_{-L}(\omega)=0$. Then the above crossing spectral density is reduced to
\begin{align}
{\mathcal{J}}_{+-R}(\tilde{V})=&-2\dfrac{\mathcal{J}_{+R}(\tilde{V})\mathcal{J}_{-R}(\tilde{V})}{\mathcal{J}_{+}(\tilde{V})\!+\!\mathcal{J}_{-}(\tilde{V})}.
\end{align}

In the  topological phase ($\lambda<1$), the inverse Bogoliubov transformation coefficients of the right Majorana zero modes  $\kappa_{0R}$ and $\kappa^\prime_{0R}$ have the same magnitude with opposite sign, see Eq.~(\ref{coupling}). 
As a result, 
we have $\mathcal{J}_{+R}(\omega)=0$ and $\mathcal{J}_{-R}(\omega)=J_R(\omega)=\mathcal{J}_{-}(\omega)$,
which also lead to ${\mathcal{J}_{+-R}}(\omega)=0$. Thus, the right differential conductance is reduced to
\begin{align}
\dfrac{dI_{0R}}{d(-V)}=\dfrac{2e^2}{h}\bigg[& \dfrac{J_R^2(\tilde{V})}{[\tilde{V}\!+\!\delta\omega_{-R}(\tilde{V})]^2\!+\!J_R^2(\tilde{V})}
\bigg] .
\end{align}
It has the same form as the left differential conductance given by Eq.(\ref{left cond0}). Consequently, the differential conductance at zero bias is also quantized $2e^2/h$, independent of the details of the spectral density $J_R(\omega)$.

On the other hand, Eq.~(\ref{coupling}) shows that the inverse bogoliubov transformation coefficients of the right Majorana zero modes $\kappa_{0R}$ and $\kappa^\prime_{0R}$ have different values in the  topological trivial phase ($\lambda>1$). In this case, $\mathcal{J}_{+R}(\omega)=\sqrt{1-\lambda^{-2}}J_R(\omega)=\mathcal{J}_{+}(\omega)$ and $\mathcal{J}_{-R}(\omega)=J_R(\omega) =\mathcal{J}_{-}(\omega)$. This means that both the left and right Majorana zero modes are coupled to the right lead but the coupling strengths are different, as shown in Fig.~\ref{fig2}(b). Now we have a non-vanished crossing spectral density ${\mathcal{J}_{+-R}}(\omega)=\sqrt{1-\lambda^{-2}}J_R(\omega)/(1+\sqrt{1-\lambda^{-2}})$. As a result, for $ \lambda>1$, the right differential conductance  contains an interference term, 
\begin{align}
\dfrac{dI_{0R}}{d(-V)}= & \dfrac{2e^2}{h}\dfrac{\sqrt{1-\lambda^{-2}}-1}{\sqrt{1-\lambda^{-2}}+1}\notag\\
&\times\Bigg[
\dfrac{(1-\lambda^{-2})J_R^2(\tilde{V})}{[\tilde{V}\!+\!\delta\omega_{+R}(\tilde{V})]^2\!+\!(1-\lambda^{-2})J_R^2(\tilde{V})} \notag \\
& ~~~~~ -\dfrac{J_R^2(\tilde{V})}{[\tilde{V}\!+\!\delta\omega_{-R}(\tilde{V})]^2\!+\!J_R^2(\tilde{V})}
\Bigg].  \label{intft}
\end{align}
At zero bias, the differential conductance ia zero due to the interference, and this conclusion is also independent of the value of $\lambda$ and the spectral density $J_R(\omega)$.

\subsubsection{Total differential conductance}

The total differential conductance can be simply obtained by combining the left and right differential conductance together, that is $I_0=\frac{1}{2}(I_{0L}-I_{0R})$ and
\begin{align}
\dfrac{dI_0}{dV}=\frac{1}{2}&\dfrac{d(I_{0L}-I_{0R})}{dV}\notag\\
=&\dfrac{e^2}{h}\bigg[\dfrac{\mathcal{J}_{+}(\tilde{V})}{[\tilde{V}\!+\!\delta\omega_+(\tilde{V})]^2\!+\!\mathcal{J}_{+}^2(\tilde{V})} \Big[\mathcal{J}_{+}(\tilde{V})\!-\!{\mathcal{J}_{+-}}(\tilde{V}) \Big]\notag\\
&~~+\!\dfrac{\mathcal{J}_{-}(\tilde{V})}{[\tilde{V}\!+\!\delta\omega_-(\tilde{V})]^2\!+\!\mathcal{J}_{-}^2(\tilde{V})}
 \Big[\mathcal{J}_{-}(\tilde{V})\!-\!{\mathcal{J}_{+-}}(\tilde{V}) \Big] \bigg] ,
\label{steady}
\end{align}
where
\begin{align}
{\mathcal{J}_{+-}}(\tilde{V})=&\dfrac{2\left(\!\sqrt{\mathcal{J}_{+L}(\tilde{V})\mathcal{J}_{-L}(\tilde{V})}-\!\sqrt{\mathcal{J}_{+R}(\tilde{V})\mathcal{J}_{-R}(\tilde{V})}\right)^2}{\mathcal{J}_{+}(\tilde{V})\!+\!\mathcal{J}_{-}(\tilde{V})}.
\end{align}
As we have shown that the interference terms play an important role on characterizing the topological phase transition through the differential conductance. The presence or absence of the interference will determine whether the value of zero-bias conductance is quantized with $2e^2/h$ or zero. 

In the topologically nontrivial phase ($\lambda<1$), the nonlocal wavefunction distribution of the zero-energy bogoliubon shown in Fig.~\ref{fig2}(a) causes the left (right) Majorana zero mode only coupled to the left (right) lead, and thus no interference between the two Majorana zero modes contributes to the zero-bias conductance. This corresponds to the Majorana resonance with perfect Andreev reflection, and the zero-bias peak of differential conductance is always quantized with $2e^2/h$. In contrast,  in the topologically trivial phase ($\lambda>1$), the wavefunction distribution of the zero-energy bogoliubon is localized at the right-hand-side of the superconducting chain. Therefore, both the left and right Majorana zero modes are coupled to the right lead with non-negligible coupling strengths. The interference term in conductance cancels the conductance of the electrons separately flowing through the channels of left and right Majorana zero modes at zero bias, resulting in a zero value of conductance, as shown in Fig.~\ref{fig2}(b). This perfect cancellation is a result of the particle-hole symmetry at the long superconducting chain limit. For finite size chains, the overlap of the left and right Majorana zero modes could break down this cancellation and result in non-zero conductance at zero-bias point, as we will discuss in the next section  

\section{Numerical analysis of the differential conductance for Lorentzian spectral density}
\label{sec4}
To explicitly see the difference between the Andreev bound state and the Majorana bound state manifested in the differential conductance, we take a Lorentzian spectral density of leads that has been used in various studies of molecular wires coupling to electron reservoirs \cite{Tu2008,Lorentz1,Lorentz2}. The Lorentzian spectral density of lead $\alpha$ in Eq.~(\ref{efsd}) has the following form
\begin{align}
&J_\alpha(\omega)=\frac{|\eta_{\alpha}|^{2}d_{\alpha}^2}{\omega^2+d_{\alpha}^2} 
\end{align}
Here $\eta_{\alpha}$ is the coupling constant between the zero-energy bogoliubon in the superconducting chain and the lead $\alpha$. The parameter $d_{\alpha}$ describes the widths of the spectrum, which characterizes the states in the lead $\alpha$ with energy around zero that effectively involve in the electron tunneling between the lead $\alpha$ and the zero-energy bogoliubon in the superconducting chain. 
We consider the symmetric case with $d_L=d_R=d$ in the following discussion.

\begin{figure}[t]
\includegraphics[scale=0.2]{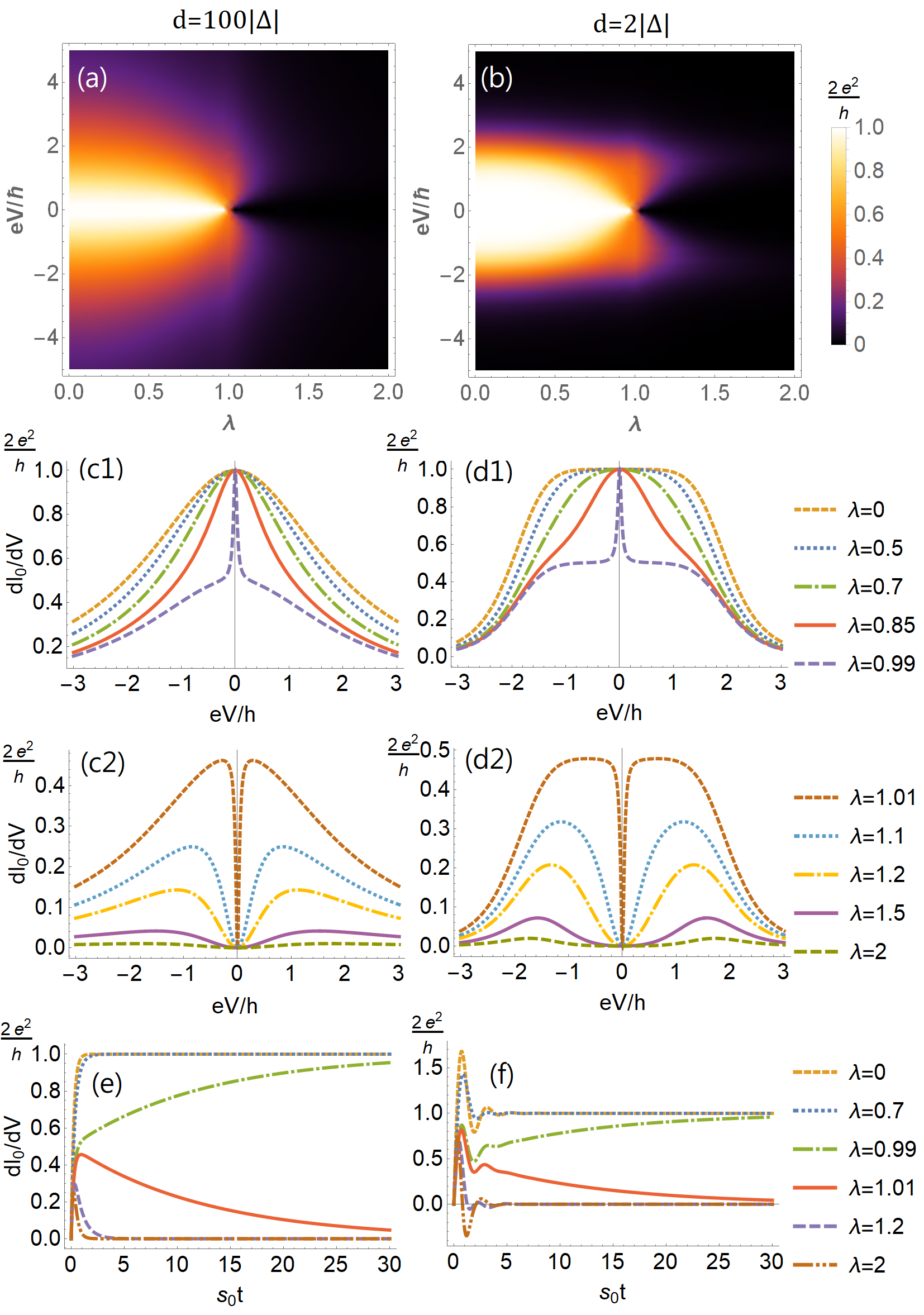}
\caption{(Colour online) (a)-(b) The contour plots of the steady-state differential conductance by varying $\lambda$ and the bias voltage $V$ for the Lorentzian spectral density width $d=100|\Delta|$ (the left panel) and $d=|\Delta|$ (the right panel). (c)-(d) The vertical line-cuts of Fig.~\ref{fig3}(a) and (b), respectively, at different values of $\lambda$. (e)-(f) The corresponding time evolution of the differential conductance at zero-bias with different values of $\lambda$, where $s_0=2|\Delta|/\hbar$. }
\label{fig3}
\end{figure}

\subsection{Topological phase transition in differential conductance}
Once the spectral density is specified, the differential conductance can be directly obtained from Eq.~(\ref{steady}). We first consider the case with $|\eta_R|=|\eta_L|=|\Delta|$. The steady-state differential conductance as a function of bias voltage $\tilde{V}$ and the rate $\lambda$ is shown in Fig.~\ref{fig3}(a) and (b), respectively, with two different spectral density widths $d=100|\Delta|$ and $d=2|\Delta|$. The spectral density approaches to the wide-band limit for the large width $d$ where the dynamic tunneling process is Markovian. While, for the small width $d$, the electron transport dynamics is non-Markovian, as it is well-known in open quantum systems \cite{PRL2012}.

The numerical results shows that the differential conductance undergoes a significant phase transition around the critical point $\lambda_c=1$, as shown in Fig.~\ref{fig3}(c) and (d),  for both the Markovian and non-Markovian transport. A dramatical change of zero-bias peak is shown in Fig.~\ref{fig3}(c) and (d), which is the vertical line-cuts of Fig.~\ref{fig3}(a) and (b) at different values of $\lambda$, respectively. Figure~\ref{fig3}(c1) and (d1) show that in the topologically nontrivial phase $\lambda<1$, the increase of $\lambda$ will narrow the width of the differential conductance versus the bias voltage, while the height is always $2e^2/h$. When we change $\lambda$ to the critical point ($\lambda=0.99$), a rather narrow peak is formed at zero bias voltage. Further increasing $\lambda$ turns the superconducting chain into the topologically trivial phase $\lambda>1$. Slightly above the critical point ($\lambda=1.01$), the patterns of differential conductance is almost the same as the result slightly below the critical point, except for the value at zero bias voltage, where a narrow valley reaching zero is formed instead of a narrow peak,  see Fig.~\ref{fig3}(c2) and (d2). This dramatical change at zero bias manifests the significant different contributions to the differential conductance through the Majorana bound state and the Andreev bound state. The latter is strongly influenced by the interference effect shown in Eq.~(\ref{intft}), while the former does not be affected due to its non-local property, as also intuitively shown by Fig.~\ref{fig2}. Also, comparison Figs.~\ref{fig3}(c1)-(d1) with Figs.~\ref{fig3}(c2)-(d2), the quantum phase transition passing through the critical point $\lambda=1$ is manifested.

Furthermore, we study the time evolution of the conductance (the differential conductance at zero bias voltage) with different values of $\lambda$, and the results are presented in Fig.~\ref{fig3}(e) and (f). It shows the different process of the formation of the zero-bias peaks in the differential conductance, as a manifestation of Markovian and non-Markovian electron transport dynamics. For $d=100|\Delta|$, the transport dynamics are Markovian. The corresponding exponential relaxation is shown in Fig.~\ref{fig3}(e). While for $d=2|\Delta|$, the oscillation is manifested in the beginning and then non-exponentially approaches to steady states. This is a typical non-Markovian process. More importantly, in both cases, the conductances with different $\lambda$ can only reach one of the two values in the long-time limit, depending on their topological property. They approach to $2e^2/h$ in the topologically nontrivial phase $\lambda<1$, or become zero in the topologically trivial phase $\lambda>1$. In fact, around the critical point ($\lambda=0.99$ and $\lambda=1.01$), the dynamics of the differential conductance in the two different phases are very similar in the beginning. After a short time, the electron interference effect is involved in the topologically trivial phase but not in the topologically nontrivial phase. This dynamically makes the differential conductance decay to zero in the topologically trivial phase but approach to $2e^2/h$ in the topological nontrivial phase. This results in an abrupt and discontinuous transition of the differential conductance is clearly manifested in the critical region around $\lambda_c=0$.

\begin{figure}[t]
\includegraphics[scale=0.26]{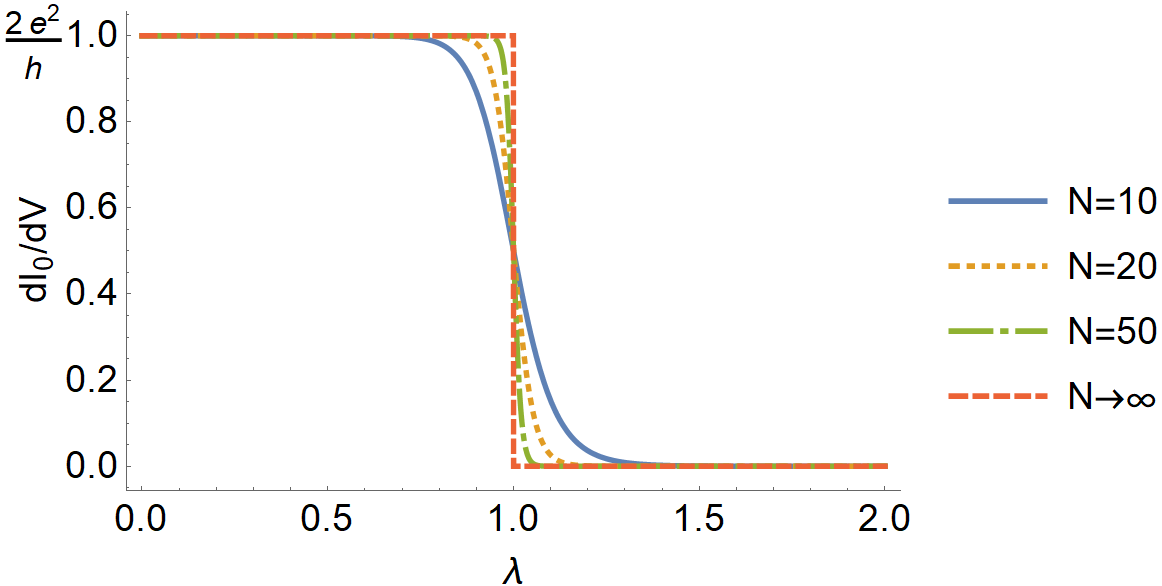}
\caption{(Colour online) The zero-bias conductance as a function of $\lambda$ for different length of the superconducting chain.}
\label{fig4}
\end{figure}

Moreover, to see the finite size effect of the superconducting chain, we also plot the differential conductance for the different lengths of superconducting chain in Fig.~\ref{fig4}. It shows that the topological phase transition process is relatively smooth for a finite length of superconducting chain, rather than the dramatical transition in the infinite length limit. In other words, in a more practical situation with a finite length, the height of the zero-bias peak in the topologically non-trivial phase $\lambda<1$ near the critical point will no longer always be $2e^2/h$ but will be reduced, and that in the topological trivial phase $\lambda>1$ will also not always be zero but could be increased. In other words, when the length of the superconducting chain is shortened, the phase transition becomes smoother.

\subsection{Effect of variation in the width and coupling strength in spectral density}

\begin{figure}[t]
\includegraphics[scale=0.3]{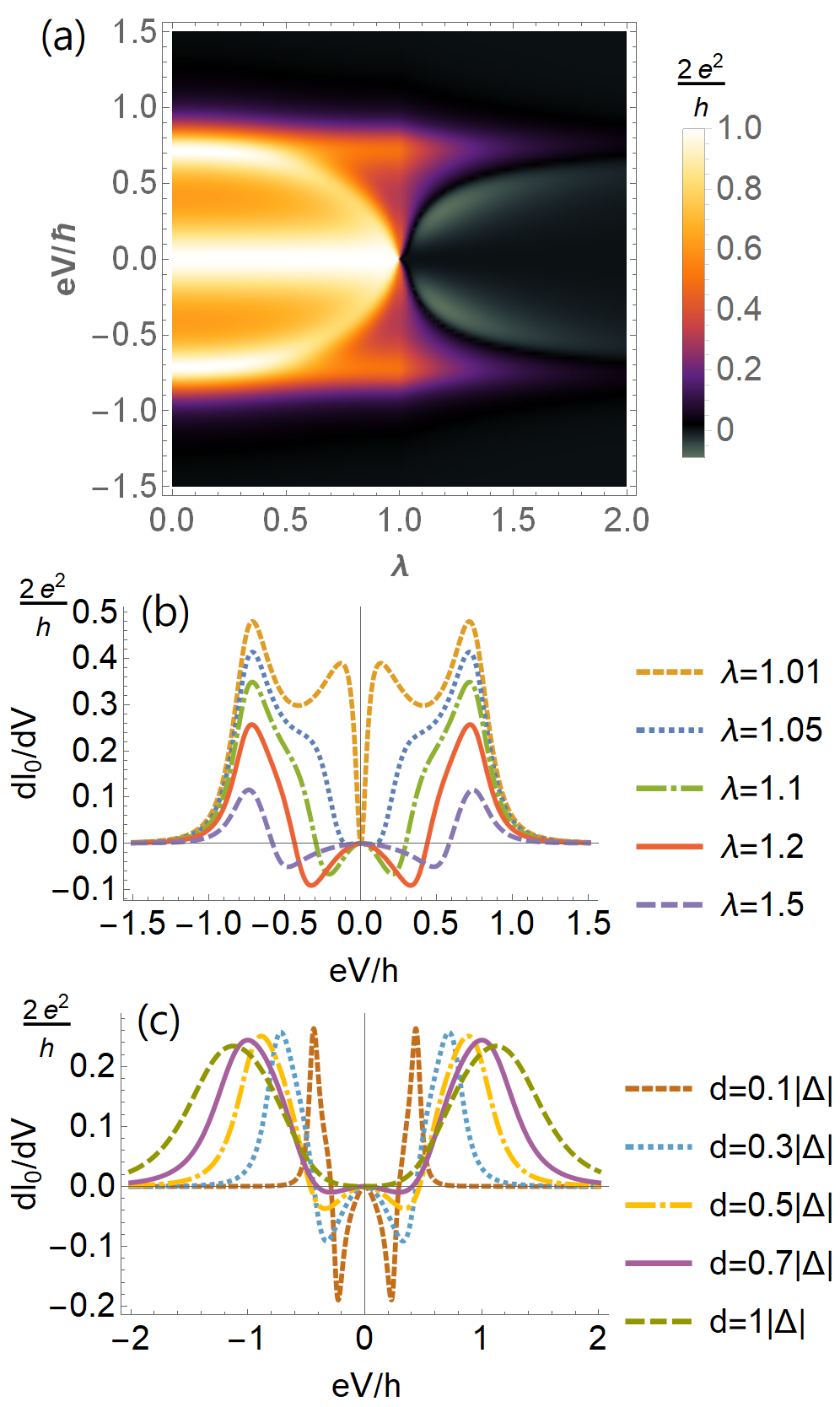}
\caption{(Colour online) (a) The steady-state differential conductance by varying $\lambda$ and the bias voltage $V$ for the Lorentzian spectral density width $d=0.3|\Delta|$. (b) The vertical line-cuts of Fig.~\ref{fig5}(a) at different values of $\lambda$. (c) The steady-state differential conductance in the topologically trivial phase $\lambda=1.2$ with different values of the spectral density width $d$.}
\label{fig5}
\end{figure}

\begin{figure}[t]
\includegraphics[scale=0.3]{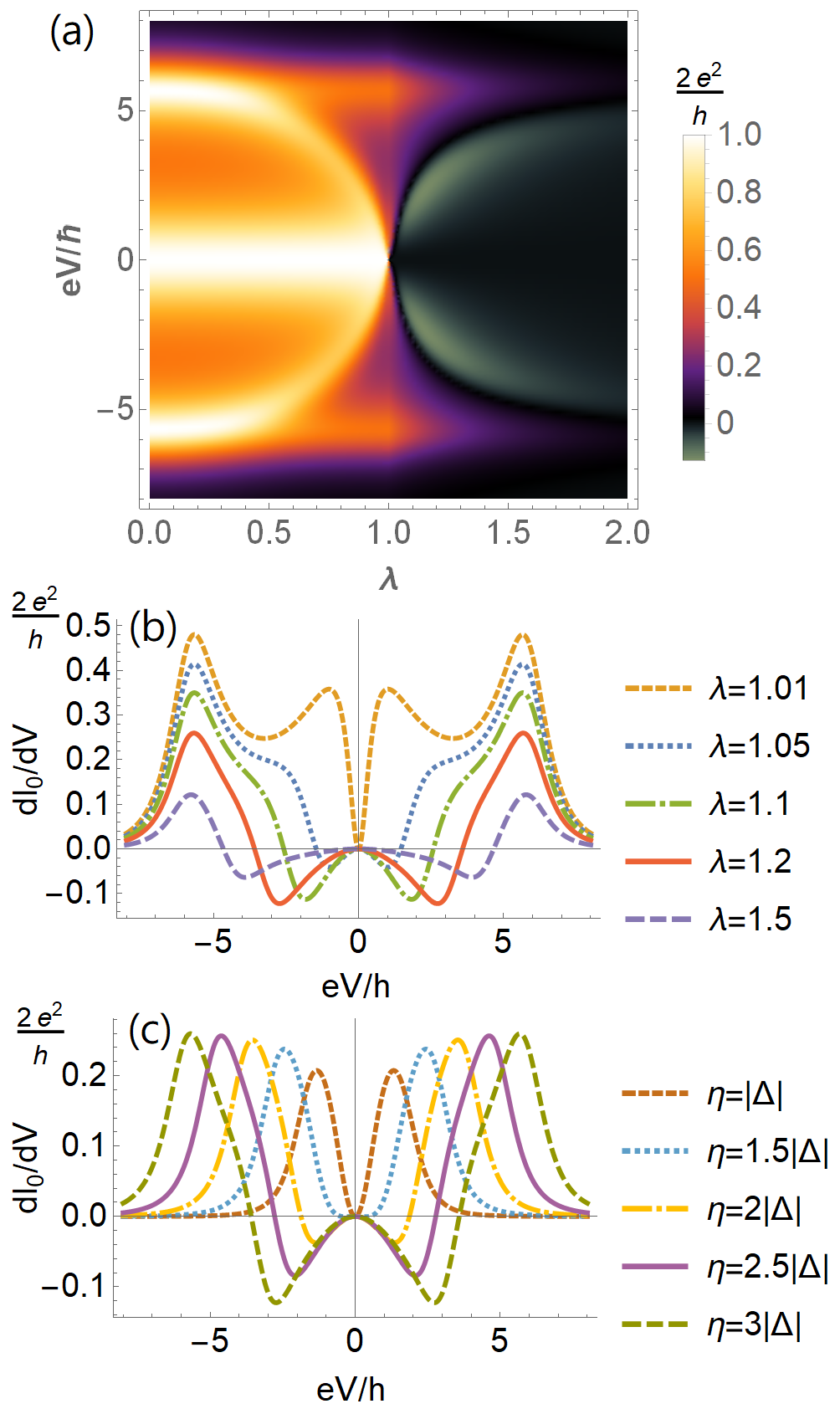}
\caption{(Colour online) (a) The steady-state differential conductance by varying $\lambda$ and the bias voltage $V$ for the Lorentzian spectral density width $d=2|\Delta|$ and the coupling strength $|\eta_L|=|\eta_R|=3|\Delta|$. (b) The vertical line-cuts of Fig.~\ref{fig6}(a) at different values of $\lambda$. (c) The steady-state differential conductance in the topologically trivial phase $\lambda=1.2$ for the spectral density width $d=2|\Delta|$ with different values of coupling strength $\eta$.}
\label{fig6}
\end{figure}

We also find some interesting results in the topologically trivial phase for the cases with a narrower spectral density or a strong coupling strength. In realistic junction devices, the spectra width $d_\alpha$ strongly depends on the part connecting the lead to the hybrid nanowire of superconductor-semiconductor systems, where the semiconductor nanowire is not covered by the superconductor \cite{Lutchyn18,QD-hybrid,Majorana nanowire3,Large ZBP}. Between the normal lead and the semiconductor nanowire, a Schottky barrier could be formed \cite{hybrid num1,PS-Andreev1}. This leads to the formation of an effective quantum dot at the end of the hybrid nanowire, which is usually be treated as an inhomogeneous energy potential in the nanowire in theoretical models \cite{hybrid num1,hybrid num2,PS-Andreev1,PS-Andreev2, barrier1, barrier2, disorder}. In this situation, the effective spectral width could be very narrow. Correspondingly, the coupling could also be strong. Note that our previous work \cite{Tu2008} has shown that when the spectral width is narrow or the coupling strength is strong, the non-Markovian processes dominate the dynamics where the backreaction memory effect plays an important role. That is, the backflow
of charges and information from the system (superconducting chain) to the environment (leads) is non-negligible.

We plot the differential conductance for a very narrow width $d=0.3|\Delta|$ of the spectral density versus bias voltage $\tilde{V}$ and $\lambda$ in Fig.~\ref{fig5}(a) and its vertical line-cuts in Fig.~\ref{fig5}(b). The results show more ups and downs in the differential conductance versus bias voltage. On the other hand, the differential conductance in topologically trivial phase ($\lambda>1$) has negative values in the low bias region. 
This phenomenon has indeed been observed in the electron tunnelling through systems with a strongly non-monotonic density of states \cite{mole1,mole2,mole3,mole4,mole5,hete1,hete2,hete3,hete4,hete5}. We also plot the differential conductance versus bias voltage with different values of spectral density width $d$ in Fig.~\ref{fig5}(c). It shows that as $d$ become smaller, the width of the valley of negative differential conductance become narrower, but the depth of the valley will also become deeper. 
In Fig.~\ref{fig6}(a) and (b), we plot further the cases for strong coupling strength $|\eta_L|=|\eta_R|=3|\Delta|$ between the superconducting chain and leads with the spectral density width $d=2|\Delta|$. The results show that there is also the valley of negative differential conductance in the low bias voltage region in topologically trivial phase. In Fig.~\ref{fig6}(c), one can find that the increase in coupling strength $\eta$ will not only widen the width of the valley of negative conductance, but also deepen its depth.

\subsection{Effect of the coherence between zero-energy and nonzero-energy bogoliubons}
In the previous discussion, due to the relatively large energy gap between the zero-energy mode and the non-zero energy bulk band in the non-critical region, we considered only the quantum transport through the zero-energy channel. The transport through the nonzero-energy channels is ignored. However, from Fig.~\ref{fig1}(a) we can find that the nonzero-energy bulk band and the zero-energy mode are very close around the critical point $\lambda_c=1$. Therefore, the coherence between zero-energy ground state and other excited states is not negligible in the critical region. 

\begin{figure}[t]
\includegraphics[scale=0.29]{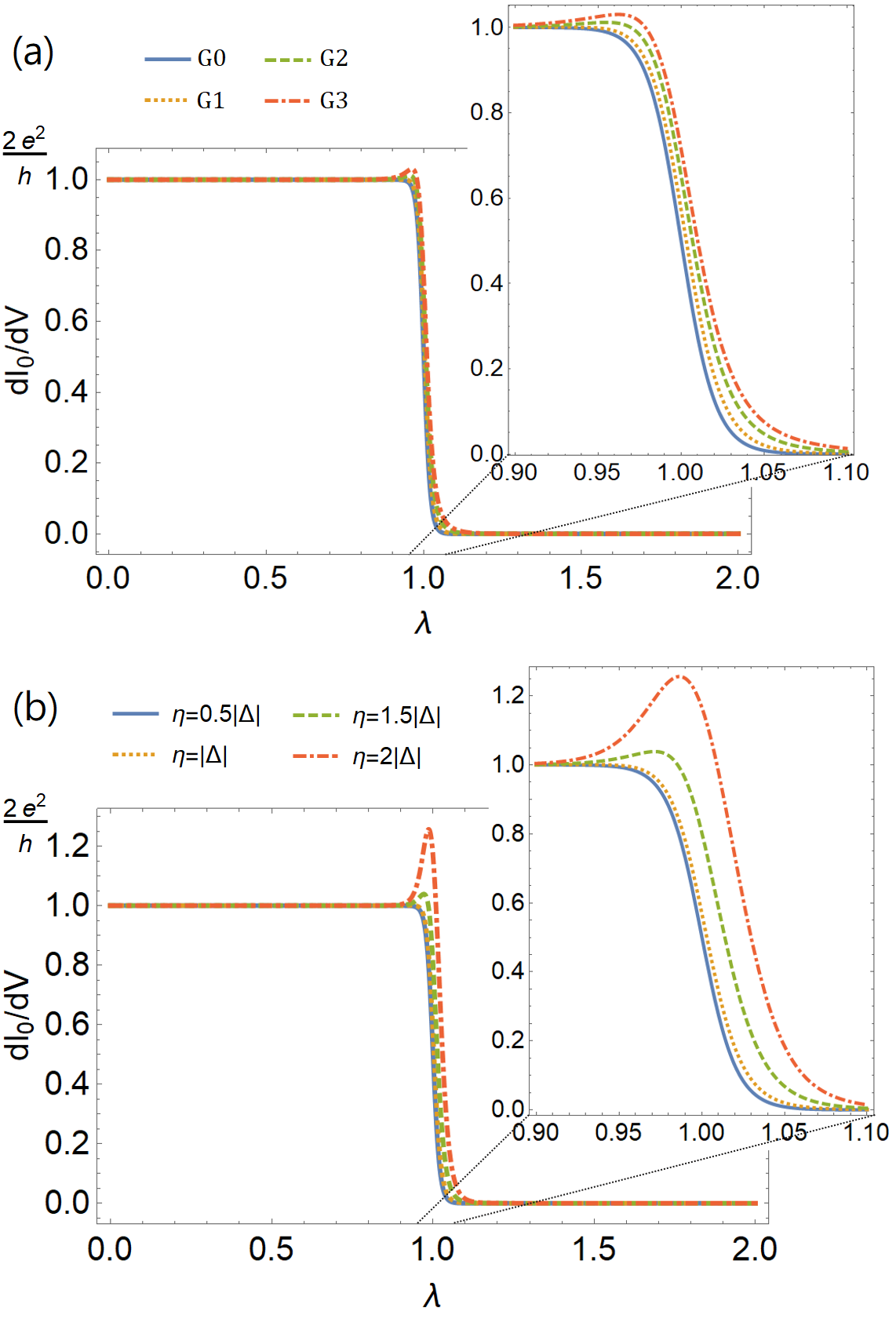}
\caption{(Colour online) The zero-bias conductance as a function of $\lambda$ for (a) the non-zero energy excited state contributions,  including the first excited state ($G1$), also the second excited state ($G2$), and also the third excited state ($G3$), comparing with only the zero-energy ground state ($G0$); (b) different coupling strength $\eta$ with the zero-energy state and the first excited state being included.}
\label{fig7}
\end{figure}

To see the effect of this coherence between zero-energy and nonzero-energy channels, we calculate the zero-bias conductance including the first few nonzero-energy bogoliubon states in Fig.~\ref{fig7} (a). We can find that the coherence between the zero-energy state and these non-zero energy states  increases the zero-bias conductance near the critical point. The more nonzero-energy bogoliubon states are taken into account, the zero-bias conductance near the critical point becomes higher. It shows that the height of zero-bias peak near the critical point can exceed the value of $2e^2/h$, even if the superconducting chain is in the topologically non-trivial phase ($\lambda<1$). 
We further study the influence of coupling strength on this coherence effect. In Fig.~\ref{fig7}(b), we show the zero-bias conductance including the zero-energy ground state and the first excited state contributions for different coupling strength. We  find that when the coupling strength becomes stronger, the nonzero-energy state enhances significantly the zero-bias conductance  near the critical point. In the strong coupling region, the zero-bias conductance peak can largely exceed $2e^2/h$. These results are similar to that observed in recent experiments and considered to be caused by disorder-induced subgap states \cite{disorder,Large ZBP,Das Sarma2021}.

\section{Conclusion}
\label{sec6}
In conclusion, using the quantum transport theory based on quantum Langevin equation approach \cite{PY2015,HL,Huang2020,01}, we analytically solve the differential conductance for this asymmetric superconductor two-terminal device with general dissipative spectral densities. This asymmetric superconductor two-terminal device contains zero-energy modes that can undergo a topological phase transition from the topologically nontrivial Majorana bound state to the topologically trivial Andreev bound state. We study the different transport properties through the Majorana bound states and Andreev bound states unambiguously in this two-terminal device.  
We show explicitly that in the steady-state limit, the differential conductance is fully determined by the bound-state-wavefunction-dependent spectral densities. In the topologically nontrivial state $\lambda<1$, the left Majorana zero mode and the right Majorana zero mode are respectively coupled to the left and right leads, and the zero-bias differential conductance is perfectly quantized with $2e^2/h$ for the Majorana bound state. In the topologically trivial state $\lambda>1$, both the left and right Majorana zero modes are localized at the right-hand side of the superconducting chain to form the zero-energy Andreev bound state, and the interference between them resulted in a zero value for the differential conductance at zero bias voltage.  
This result suggests that such a Majorana device can be considered as an ideal quantum diode. 

We numerically clarify the formation of zero-bias conductance peak in the non-Markovian transport process. We also verify that for the ideal case with a long superconducting chain length and negligible coherence between zero-energy and nonzero-energy bogoliubons, the topological phase transition can be  manifested in the dramatic change of zero-bias conductance peak from the value of $2e^2/h$ to zero. This is independent of the shape of spectral density and the coupling between the superconducting chain and leads. We find the negative differential conductance in topological trivial phase if the zero modes coupled to a narrow band of the leads, which is indeed a useful feature in electronic semiconductor devices as oscillators and amplifiers. Our numerical result shows that the depth of the negative differential conductance and its range sensitively depend on the width of spectral density and the coupling strength which are experimentally controllable.

We also show the finite size effect of the superconducting chain length and the nonzero-energy bogoliubon state contributions to the differential conductance in this analytically solvable two-terminal device. The significant changes of zero-bias conductance peak near the critical point of
the topological phase transition are observed. Shortening the superconducting chain length causes the zero-bias conductance peak near the critical point to decrease for the topologically nontrivial phase but increase for the topologically trivial phase. In contrast, the nonzero-energy channels could enhance the conductance near the critical point for both the topologically non-trivial and topologically trivial phases. This results in zero-bias conductance peak being able to exceed the quantized value $2e^2/h$, and possibly relating the results observed in recent experiments \cite{disorder,Large ZBP,Das Sarma2021}. We expect such an analytically solvable system with unambiguous existence of both the zero-energy Majorana bound states and zero-energy Andreev bound states could help our understanding of Majorana quasiparticle and its applications.

\acknowledgments
This work is supported by Ministry of Science and Technology of Taiwan, Republic of China under
Contract No. MOST-108-2112-M-006-009-MY3. 

\appendix
\section{}
\label{appendixA}
\begin{widetext}
In this appendix, we outline the derivation of the exact non-Markovian dynamics of the transport current and the differential conductance. The general transient transport current flowing from lead $\alpha$ into the superconducting chain is defined by
\begin{align}
I_{\alpha}(t)=-e \dfrac{d}{dt}\left\langle N_\alpha(t) \right\rangle 
= \frac{e}{i\hbar} \left\langle [H, N_\alpha(t)]\right\rangle,
\end{align}
where $N_\alpha(t)=\sum\limits_{k} b_{\alpha k}^{\dagger}(t)b_{\alpha k}(t)$ is the particle number operator in lead $\alpha$. Using the Heisenberg equation of motion with Eq.~(\ref{Hb}), one has
\begin{align}
b_{\alpha k}(t)=e^{-\frac{i}{\hbar}\epsilon_{\alpha k}(t-t_{0})}b_{\alpha k}(t_{0})+\frac{i}{\hbar}\sum\limits_j \int_{t_{0}}^{t} e^{-i\epsilon_{\alpha k}(t-\tau)}\eta_{\alpha k}\big[\kappa_{\alpha j}a_{j}(\tau)+\kappa_{\alpha j}^{\prime}a^{\dagger}_{j}(\tau)\big]d\tau .
\label{EOMb}
\end{align}
Therefore, the equation of transport current becomes
\begin{align}
I_{\alpha}(t)=&\sum\limits_{k}\dfrac{1}{\hbar^2}\int \bigg\langle e^{-\frac{i}{\hbar}\epsilon_{\alpha k}(t-\tau)}[\kappa^*_{i\alpha k}a_{i}^{\dagger}(t)+\kappa_{i\alpha k}^{\prime*}a_{i}(t)][\kappa_{i\alpha k}a_{i}(\tau)+\kappa_{i\alpha k}^{\prime}a_{i}^{\dagger}(\tau)]+H.c. \bigg\rangle d\tau\notag\\
&+\sum\limits_{k}\dfrac{i}{\hbar}\bigg\langle e^{-\frac{i}{\hbar}\epsilon_{\alpha k}(t-t_{0})}[\kappa^*_{i\alpha k}a_{i}^{\dagger}(t)b_{\alpha k,i}(t_{0})+\kappa_{i\alpha k}^{\prime*}a_{i}(t)b_{\alpha k,i}(t_{0})]+H.c. \bigg\rangle,
\label{I1}
\end{align}

Furthermore, the Heisenberg equations of motion for the bogoliubon modes $\boldsymbol{a}(t) \equiv (a_0(t), a_1(t), a_2(t), \cdots, a_N(t))^T$ and 
$\boldsymbol{a^{\dagger}}(t)=(a^\dag_0(t), a^\dag_1(t), a^\dag_2(t), \cdots, a^\dag_N(t))^T$, after eliminating the degrees of freedom of the two leads, becomes
\begin{align}
\frac{d}{dt}\!\left(\!\!\begin{array}{c}
\boldsymbol{a}(t)\\
\boldsymbol{a^{\dagger}}(t)
\end{array}\!\!\right)
&\!+\!\frac{i}{\hbar}\!\left(\!\!\begin{array}{cc}
\boldsymbol{\varepsilon}
&0\\
0
&-\boldsymbol{\varepsilon}
\end{array}\!\!\right)\!\!
\left(\!\!\begin{array}{c}
\boldsymbol{a}(t)\\
\boldsymbol{a^{\dagger}}(t)
\end{array}\!\!\right) \!+ \!\!\int_{t_{0}}^{t} \!\!\!
\boldsymbol{G}(t,\tau)\!
\left(\!\!\begin{array}{c}
\boldsymbol{a}(\tau)\\
\boldsymbol{a^{\dagger}}(\tau)
\end{array}\!\!\right)d\tau
=\left(\!\!\begin{array}{c}
\boldsymbol{\xi}(t)\\
\boldsymbol{\xi^{\dagger}}(t)
\end{array}\!\!\right)
\label{EOMa}
\end{align}
which is a generalized quantum Langevin equation \cite{PY2015,HL,01,Huang2020}. The third term in the left side of Eq.~(\ref{EOMa}) corresponds to the damping, and the right-hand side of the equation is the quantum noise force. The quantum noise force is given by
\begin{align}
\left(\!\!\begin{array}{c}
\boldsymbol{\xi}(t)\\
\boldsymbol{\xi^{\dagger}}(t)
\end{array}\!\!\right)=\frac{i}{\hbar}\sum\limits_{\alpha k} \!\left(\!\!\begin{array}{cc}
\eta_{\alpha k}^*\boldsymbol{\kappa^*}_{\alpha}e^{-\frac{i}{\hbar}\epsilon _{\alpha k}\tau}
&-\eta_{\alpha k}\boldsymbol{\kappa^{\prime}}_{\alpha}e^{\frac{i}{\hbar}\epsilon _{\alpha k}\tau}\\
\eta_{\alpha k}^*\boldsymbol{\kappa^{\prime*}}_{\alpha }e^{-\frac{i}{\hbar}\epsilon _{\alpha k}\tau}
&-\eta_{\alpha k}\boldsymbol{\kappa}_{\alpha}e^{\frac{i}{\hbar}\epsilon _{\alpha k}\tau}
\end{array}\!\! \right)\!\!
\left(\!\!\begin{array}{c}
b_{\alpha k}(t_{0})\\
b_{\alpha k}^{\dagger}(t_{0})
\end{array}\!\!\right),
\end{align}
which is associated with the initial states of leads.

Due to the linearity of Eq.~(\ref{EOMa}), its general solution has the form as \cite{Tu2008,Tu2010,PY2015,HL,01}
\begin{align}
\left(\!\!\begin{array}{c}
\boldsymbol{a}(t)\\
\boldsymbol{a^{\dagger}}(t)
\end{array}\!\!\right)=\boldsymbol{U}&(t,t_{0})\left(\!\!\begin{array}{c}
\boldsymbol{a}(t_{0})\\
\boldsymbol{a^{\dagger}}(t_{0})
\end{array}\!\!\right) +
\left(\!\!\begin{array}{c}
\boldsymbol{f}(t,t_0)\\
\boldsymbol{f^{\dagger}}(t,t_0)
\end{array}\!\!\right) .
\label{EOM}
\end{align}
Here, 
\begin{align}
\boldsymbol{U}(t,t_{0})=\left(\!\! \begin{array}{cc}
\langle\{\boldsymbol{a}(t),\boldsymbol{a^{\dagger}}(t_{0})\}\rangle & \langle\{\boldsymbol{a}(t),\boldsymbol{a}(t_{0})\}\rangle\\
\langle\{\boldsymbol{a^{\dagger}}(t),\boldsymbol{a^{\dagger}}(t_{0})\}\rangle & \langle\{\boldsymbol{a^{\dagger}}(t),\boldsymbol{a}(t_{0})\}\rangle
\end{array} \!\!\right)
\end{align}
is an generalization of the usual nonequilibrium retarded Green function to incorporate with pairings. It obeys the generalized 
Dyson equation given by Eq.~(\ref{U}). By applying the modified Laplace transformation $\tilde{\boldsymbol{U}}(s)=\int_{t_{0}}^{\infty}\boldsymbol{U}(t,t_{0})e^{is(t-t_{0})}$ introduced by our previous work \cite{01}, it becomes
\begin{align}
&\tilde{\boldsymbol{U}}(s)=i\left( \begin{array}{cc}
s-\boldsymbol{\tilde{\epsilon}}-\boldsymbol{\Sigma}(s)-\boldsymbol{\hat{\Sigma}}(-s) & \boldsymbol{\bar{\Sigma}}(s)+\boldsymbol{\bar{\Sigma}}(-s)\\
\boldsymbol{\bar{\Sigma}}(s)+\boldsymbol{\bar{\Sigma}}(-s) & s+\boldsymbol{\tilde{\epsilon}}-\boldsymbol{\Sigma}(-s)-\boldsymbol{\hat{\Sigma}}(s)
\end{array}\right)^{-1},
\label{LaplaceU}
\end{align}
where $\boldsymbol{\tilde{\epsilon}}=\boldsymbol{\epsilon}/\hbar$ and the self-energy corrections $\boldsymbol{\Sigma}(s)$, $\boldsymbol{\bar{\Sigma}}(s)$, and $\boldsymbol{\hat{\Sigma}}(s)$ are the Laplace transform of the matrix elements in Eq.~(\ref{G0})
\begin{align}
\boldsymbol{\Sigma}(s)=\int\dfrac{d\omega}{2\pi}\dfrac{\boldsymbol{J}(\omega)}{s-\omega}\xrightarrow{s=\omega\pm i0^{+}}\boldsymbol{\delta\omega}(\omega)\mp\dfrac{i}{2}\boldsymbol{J}(\omega),
\label{sigma}
\end{align}

Then by applying the inverse transformation to Eq.~(\ref{LaplaceU}), we can analytically solve $\boldsymbol{U}(t,t_{0})$, which consists of a summation of dissipationless oscillations arose from localized modes (localized bound states) determined by the real part of the self-energy correction to the probing spin, plus nonexponential decays induced by the discontinuity of the imaginary part of the self-energy correction cross the real axes in the complex plane \cite{PRL2012}
\begin{align}
\boldsymbol{U}(t,t_{0})=\sum\limits_{s_{p}}\left(\begin{array}{cc}
\boldsymbol{X}(s_{p})&\boldsymbol{\bar{X}}(s_{p})\\
\boldsymbol{\bar{X}}(s_{p})&\boldsymbol{X}(-s_{p})
\end{array}\right)e^{-is_{p}(t-t_{0})}+\int_{-\infty}^{\infty}\dfrac{ds}{2\pi}
\left(\begin{array}{cc}
\boldsymbol{Y}(s)&\boldsymbol{\bar{Y}}(s)\\
\boldsymbol{\bar{Y}}(s)&\boldsymbol{Y}(-s)
\end{array}\right)e^{-is(t-t_{0})},
\label{exactU}
\end{align}
where $\{s_{p}\}$ is the set of the poles for the determinant of $\tilde{\boldsymbol{U}}(s)$ located at the real axis. Specifically, for special cases where the coherence between each mode is negligible, the matrix elements in Eq.~(\ref{exactU}) can be solved separately for each mode $k$
\begin{subequations}
\begin{align}
&X_k(s)=\dfrac{(s+\tilde{\epsilon}_{k}-\Delta_{d,k}(s))^{2}}{[(s+\tilde{\epsilon}_{k}-\Delta_{d,k}(s))^{2}+\Delta_{o,k}^{2}(s)](\Delta_{d,k}^{\prime}(s)-1)+2\Delta_{o,k}(s)(s+\tilde{\epsilon}_{k}-\Delta_{d,k}(s))\Delta_{o,k}^{\prime}(s)}\\
&\bar{X}_k(s)=\dfrac{\Delta_{o,k}(s)^{2}}{[(s-\Delta_{d,k}(s))^{2}+\Delta_{o,k}^{2}(s)-\tilde{\epsilon}_{k}^{2}]\Delta_{o,k}^{\prime}(s)+2\Delta_{o,k}(s)(s-\Delta_{d,k}(s))(\Delta_{d,k}^{\prime}(s)-1)}\\
&Y_k(s)=\dfrac{J_{d,k}(s)[(s+\tilde{\epsilon}_{k}-\Delta_{d,k}(s))^{2}+\Delta_{o,k}^{2}(s)+\frac{J_{d,k}^{2}(s)-J_{o,k}^{2}(s)}{4}]+2J_{o,k}(s)\Delta_{o,k}(s)(s+\tilde{\epsilon}_{k}-\Delta_{d,k}(s))}{[(s-\Delta_{d,k}(s))^{2}-\tilde{\epsilon}_{k}^{2}-\Delta_{o,k}^{2}(s)-\frac{J_{d,k}^{2}(s)-J_{o,k}^{2}(s)}{4}]^{2}+[(s-\Delta_{d,k}(s))J_{d,k}(s)+\Delta_{o,k}(s)J_{o,k}(s)]^{2}}\\
&\bar{Y}_k(s)=-\dfrac{J_{o,k}(s)[(s-\Delta_{d,k}(s))^{2}-\tilde{\epsilon}_{k}^{2}+\Delta_{o,k}^{2}(s)-\frac{J_{d,k}^{2}(s)-J_{o,k}^{2}(s)}{4}]+2J_{d,k}(s)\Delta_{o,k}(s)(s-\Delta_{d,k}(s))}{[(s-\Delta_{d,k}(s))^{2}-\tilde{\epsilon}_{k}^{2}-\Delta_{o,k}^{2}(s)-\frac{J_{d,k}^{2}(s)-J_{o,k}^{2}(s)}{4}]^{2}+[(s-\Delta_{d,k}(s))J_{d,k}(s)+\Delta_{o,k}(s)J_{o,k}(s)]^{2}},
\end{align}
\end{subequations}
where $\Delta_{d,k}(s)=\boldsymbol{\delta\omega}_k(s)-\boldsymbol{\hat{\delta}}_k\omega(-s)$, $\Delta_{o,k}(s)=\boldsymbol{\bar{\delta}\omega}_k(s)-\boldsymbol{\bar{\delta}\omega}_k(-s)$, $J_{d,k}(s)=\boldsymbol{J}_k(s)+\boldsymbol{\hat{J}}_k(-s)$, and $J_{o,k}(s)=\boldsymbol{\bar{J}}_k(s)+\boldsymbol{\bar{J}}_k(-s)$. In particular, for a symmetric spectral density $J(s)=J(-s)$, these matrix elements of zero energy mode $\tilde{\epsilon}_{0}$ can be reduced to
\begin{subequations}
\begin{align}
X_0(s)=&\dfrac{1}{2}\left[\dfrac{1}{1-\delta\omega_+(s)}+\dfrac{1}{1-\delta\omega_-(s)}\right]\\
\bar{X}_0(s)=&\dfrac{1}{2}\left[\dfrac{1}{1-\delta\omega_+(s)}-\dfrac{1}{1-\delta\omega_-(s)}\right]\\
Y_0(s)=&\dfrac{\mathcal{J}_+(s)}{[s-\delta\omega_+(s)]^2+\mathcal{J}^2_+(s)}+\dfrac{\mathcal{J}_-(s)}{[s-\delta\omega_-(s)]^2+\mathcal{J}^2_-(s)}\\
\bar{Y}_0(s)=&\dfrac{\mathcal{J}_+(s)}{[s-\delta\omega_+(s)]^2+\mathcal{J}^2_+(s)}-\dfrac{\mathcal{J}_-(s)}{[s-\delta\omega_-(s)]^2+\mathcal{J}^2_-(s)},
\end{align}
\end{subequations}
where $\mathcal{J}_+(s)=J_{d,0}(s)+J_{o,0}(s)$, $\mathcal{J}_-(s)=J_{d,0}(s)-J_{o,0}(s)$.

On the other hand, the function $\{\boldsymbol{f}(t,t_{0})\}$ in Eq.~(\ref{EOM}) is the noise source characterizing the dynamics of the noise forces. Its general solution is given by 
\begin{align}
&\left(\!\!\begin{array}{c}
\boldsymbol{f}(t,t_0)\\
\boldsymbol{f^{\dagger}}(t,t_0)
\end{array}\!\!\right) =  \! \int_{t_{0}}^{t} \!\!\! d\tau \boldsymbol{U}(\tau,t_{0})
\left(\!\!\begin{array}{c}
\boldsymbol{\xi}(\tau)\\
\boldsymbol{\xi^{\dagger}}(\tau)
\end{array}\!\!\right).
\end{align}
From this solution, we obtain the generalized nonequilibrium correlation Green function
\begin{align}
\boldsymbol{V}(\tau, t) =& \, \Big\langle \left(\!\!\begin{array}{c} \boldsymbol{f^{\dagger}}(\tau,t_0)\\ \boldsymbol{f}(\tau,t_0) \end{array}\!\!\right)\!\Big(\!\!\begin{array}{cc} \boldsymbol{f}(t,t_0) & \! \boldsymbol{f^{\dagger}}(t,t_0) \end{array}\!\!\Big)
\Big\rangle,
\end{align}
whose solution is given by Eq.~(\ref{V}). Substituting these results into Eq.~(\ref{I1}), it can be simplified into the form in Eq.~(\ref{I}). 

By taking further the derivative of the transport current with respect to the bias voltage, we obtain the differential conductance. In the case of symmetric spectral density $J(s)=J(-s)$, we have
\begin{align}
\dfrac{dI_\alpha(t)}{dV}=&\dfrac{e^2}{h}\operatorname{Re}\, \Tr\bigg[
\int_{t_{0}}^t d\tau\int d\omega\dfrac{\beta/2}{1+\cosh[\beta(\omega-\tilde{V})]}[\boldsymbol{\mathbb{J}_{\alpha}}(\omega)+\boldsymbol{\hat{\mathbb{J}}_{\alpha}}(\omega)]e^{i\tilde{V}(t-\tau)}\boldsymbol{U}(\tau,t_0)\notag\\
&-\int_{t_0}^t d\tau\int \dfrac{d\omega}{2\pi} [\boldsymbol{\mathbb{J}_{M}}(\omega)+\boldsymbol{\hat{\mathbb{J}}_{M}}(\omega)]e^{-i\omega(t-\tau)}\int_{t_0}^\tau d\tau_1\int_{t_0}^t d\tau_2\int d\omega^\prime \boldsymbol{U}(\tau,\tau_1)\notag\\
&~~~~~\times\dfrac{\beta/2}{1+\cosh[\beta(\omega^\prime-\tilde{V})]}[\boldsymbol{\mathbb{J}_{M}}(\omega^\prime)+\boldsymbol{\hat{\mathbb{J}}_{M}}(\omega^\prime)]e^{-i\tilde{V}(\tau_1-\tau_2)}\boldsymbol{U^\dagger}(t,\tau_2)
\bigg],
\label{condT}
\end{align}
where $\boldsymbol{\mathbb{J}_{\alpha}}(\omega)$, $\boldsymbol{\hat{\mathbb{J}}_{\alpha}}(\omega)$, $\boldsymbol{\mathbb{J}_{M}}(\omega)$, and $\boldsymbol{\hat{\mathbb{J}}_{M}}(\omega)$ are given by Eq.~(\ref{spectral_pm}).
When the leads are initially at zero temperature ($\beta\rightarrow\infty$), the frequency dependent term in Eq.~(\ref{condT}) is reduced to a delta function: $\frac{\beta/2}{1+\cosh[\beta(\omega-\tilde{V})]}\rightarrow\delta(\omega-\tilde{V})$, then Eq.~(\ref{condT}) can be reduced to
\begin{subequations}
\label{cond}
\begin{align}
\dfrac{dI_L(t)}{dV}=&\dfrac{e^2}{h}\operatorname{Re}\, \Tr\bigg[
\!\!  \int_{t_{0}}^t \!\! d\tau[\boldsymbol{\mathbb{J}_{L}}(\tilde{V})+\boldsymbol{\hat{\mathbb{J}}_{L}}(\tilde{V})]e^{i\tilde{V}(t-\tau)}\boldsymbol{U}(\tau,t_0)\notag\\
&- \int_{t_0}^t \!\!  d\tau \!\!  \int \dfrac{d\omega}{2\pi}	 [\boldsymbol{\mathbb{J}_{L}}(\omega)-\boldsymbol{\hat{\mathbb{J}}_{L}}(\omega)]e^{-i\omega(t-\tau)} \!\!  \int_{t_0}^\tau \!\!  d\tau_1 \!\!  \int_{t_0}^t \!\! d\tau_2 \boldsymbol{U}(\tau,\tau_1)[\boldsymbol{\mathbb{J}_{M}}(\tilde{V})-\boldsymbol{\hat{\mathbb{J}}_{M}}(\tilde{V})]e^{-i\tilde{V}(\tau_1-\tau_2)}\boldsymbol{U^\dagger}(t,\tau_2)
\bigg],\\
\dfrac{dI_R(t)}{dV}=&\dfrac{e^2}{h}\operatorname{Re}\, \Tr\bigg[
\!\!-  \int_{t_{0}}^t \!\! d\tau[\boldsymbol{\mathbb{J}_{R}}(\tilde{V})+\boldsymbol{\hat{\mathbb{J}}_{R}}(\tilde{V})]e^{i\tilde{V}(t-\tau)}\boldsymbol{U}(\tau,t_0)\notag\\
&- \int_{t_0}^t \!\!  d\tau \!\!  \int \dfrac{d\omega}{2\pi}	 [\boldsymbol{\mathbb{J}_{R}}(\omega)-\boldsymbol{\hat{\mathbb{J}}_{R}}(\omega)]e^{-i\omega(t-\tau)} \!\!  \int_{t_0}^\tau \!\!  d\tau_1 \!\!  \int_{t_0}^t \!\! d\tau_2 \boldsymbol{U}(\tau,\tau_1)[\boldsymbol{\mathbb{J}_{M}}(\tilde{V})-\boldsymbol{\hat{\mathbb{J}}_{M}}(\tilde{V})]e^{-i\tilde{V}(\tau_1-\tau_2)}\boldsymbol{U^\dagger}(t,\tau_2)
\bigg].
\end{align}
\end{subequations}
If we further take the steady-state limit, the integral over time in Eq.~(\ref{cond}) is simply given by modified Laplace transformation in terms of frequency $\tilde{V}=eV/\hbar$ to the Green function $\boldsymbol{U}^{\dagger}(t,t_0)$, as shown in Eq.~(\ref{LaplaceU}). As a result, the differential conductance in Eq.~(\ref{cond}) can be further reduced to Eq.(\ref{DC}).
Moreover, if only the zero-energy bogoliubon channel is considered,  the Laplace transformation of $\boldsymbol{U}^{\dagger}(t,t_0)$ in Eq.~(\ref{LaplaceU}) is reduced to
\begin{align}
\tilde{U}(\tilde{V})=i\left(\begin{array}{cc}
u_+(\tilde{V})+u_-(\tilde{V}) & u_+(\tilde{V})-u_-(\tilde{V})\\
u_+(\tilde{V})-u_-(\tilde{V}) & u_+(\tilde{V})+u_-(\tilde{V})
\end{array}\right),
\label{U0}
\end{align}
where
$u_+(\tilde{V})=[\tilde{V}-\delta\omega_+(\tilde{V})+i\mathcal{J}_+(\tilde{V})]^{-1}$ and $u_-(\tilde{V})=[\tilde{V}-\delta\omega_-(\tilde{V})+i\mathcal{J}_-(\tilde{V})]^{-1}$.
Substituting this result into Eq.(\ref{DC}), we obtain the simple relations between the spectral densities and the differential conductance, as shown by Eq.(\ref{steadyL}) and Eq.(\ref{steadyR}).
\end{widetext}

\end{document}